\documentclass[preprints,article,accept,moreauthors,pdftex]{Definitions/mdpi}

\usepackage{amssymb}
\usepackage{amsmath}
\usepackage{amsmath,amssymb,amsfonts}%

\usepackage{mathrsfs}%

\def\eps{\varepsilon}

\def\folgt{\quad\Rightarrow\quad}

\usepackage{color}
\usepackage{soul}
\usepackage{bbding}

\firstpage{1} 
\pubvolume{xx}
\issuenum{1}
\articlenumber{5}
\pubyear{2025}
\copyrightyear{2025}
\history{Received: June 16, 2026; Accepted: date; Published: date}

\Title{Thermodynamic description of wealth inequality\\
  in the world}

\Author{  Klaus~M.~Frahm $^{1}$\orcidA{}
  Leonardo~Ermann $^{2}$\orcidB{},
 and 
Dima~L.~Shepelyansky$^{1,*}$\orcidC{}}

\AuthorNames{Klaus M. Frahm, Leonardo Ermann and Dima L. Shepelyansky}

\address{%
$^{1}$ \quad Univ Toulouse, CNRS, Laboratoire de Physique Th\'eorique,  
Toulouse, France\\
$^{2}$ \quad Departamento de F\'{\i}sica Te\'orica, GIyA,
         Comisi\'on Nacional de Energ\'{\i}a At\'omica.
           Av.~del Libertador 8250, 1429 Buenos Aires, Argentina\\
}

\corres{Correspondence: dima@irsamc.ups-tlse.fr; Tel. +33-56155-60-68 (D.L.S.)}

\abstract{According to the recent Wealth Thermalization Hypothesis (WTH)
the wealth inequality in the world
is described by the Rayleigh-Jeans (RJ) thermal distribution
of interacting agents in a society with social stratification.
In this concept, the  wealth layers of society are associated with
energy levels  from a nonlinear dynamical system conserving two
integrals of motion being total energy and probability norm.
This leads to RJ condensation and the formation
of a huge poverty phase of low wealth and a tiny oligarchic phase
that captures a main part of total society wealth.
This RJ phenomenon  has similarities
with self cleaning in multimode optical fibers and
constraint driven  condensation in various physical systems.
We analyze real Lorenz and Pareto
curves  for wealth of households in countries and the world,
Gross Domestic Product of countries,
market capitalization of companies at stock exchange of
Hong Kong, Shanghai, London, bitcoin transactions,
world trade between countries
and show that the WTH theory gives
a good description of these curves.
On the basis of this comparison
we argue that the RJ thermal distribution
provides a universal description
of wealth inequality in the world.
}
  
\keyword{Rayleigh-Jeans thermalization; wealth inequality; Lorenz and Pareto curves; GDP; stock exchange markets; bitcoins}

\begin{document}

\section{Introduction}

Statistical mechanics finds  enormously broad applications in various
fields of science (see e.g. \cite{landau}). Thermodynamic  probability
distributions for many degrees of freedom describe numerous physical
systems including the black-body cosmic microwave background
in the Universe where the Bose-Einstein distribution is 
amazingly accurate with deviations less than $10^{-3}$ in relative units
(see e.g. \cite{fixsen,wikicmb}). The application of ideas and methods of
statistical mechanics to economy and finance became  known as
econophysics \cite{mantegna}. 
In this context, arguments for the Boltzmann-Gibbs distribution of 
money, wealth and income in human society, also based on real data, 
have been presented in the stimulating review on 
applications of statistical mechanics in economy 
\cite{yakovenko1,yakovenko2}. 

The striking feature of the wealth distribution of world countries
is the strong inequality of wealth of households
as reviewed e.g. in \cite{piketty1,piketty2,boston}.
Indeed, according to \cite{piketty2}
for the whole world 50\% of the population owns only 2\% of total wealth,
while 10\% of population owns 75\% of total wealth
and 1\% of population owns 38\% of total wealth.

Various approaches based on  statistical mechanics 
and physical kinetics \cite{landau,lifshitz}
have been used by different research groups
\cite{yakovenko1,yakovenko2,angle,redner,bouchaud,chakraborti,boghosian1,boghosian2}
to explain the phenomenon of wealth inequality.
Diverse models of interacting agents are investigated
including Random Asset Exchange models studied there.
In  certain of these models, 
there is the appearance of some kind of oligarchic phase with  
a significant wealth accumulation by a group of agents
\cite{bouchaud,boghosian1,boghosian2}.
Different  arguments are presented
in favor of the Boltzmann-Gibbs type description of the 
distribution of money, wealth and income in \cite{yakovenko1,yakovenko2}.
A nonlinear Fokker–Planck description of asset exchange
is pushed forward in \cite{boghosian1,boghosian2} with the emergence of 
an oligarchic phase. 
The importance of two conserved integrals of system evolution
is  stressed in \cite{boghosian1,boghosian2}. These integrals  are
the total wealth and total probability norm (or number of agents).
Furthermore, it was argued that in the small-transaction approximation
it is important to  consider wealth instead of money or income.

An  efficient description of wealth distribution
is known as the Lorenz curve
\cite{lorenz,boston} which represents the dependence of cumulated 
normalized wealth $0\leq w \leq 1$ on the cumulated normalized fraction of
population or households $0 \leq h \leq 1$.
The case of perfect equipartition of wealth corresponds to
the diagonal $w=h$ and the doubled area between diagonal
and the Lorenz curve $w(h)$ gives the Gini coefficient
$0 \leq G \leq 1$  \cite{gini,boston}.
Typical values of the Gini coefficient $G$ are presented in \cite{wikigini}
for world countries in 2021 being in the range
$0.59 < G < 0.90$; for the whole world $G = 0.889$
(this data come from the Global Wealth Databook by Credit Suisse).
High values of the Gini coefficient being close to unity correspond to
a high inequality in a country with a large fraction of poor population
and a tiny oligarchic population fraction that owns
a huge fraction of total wealth \cite{piketty1,piketty2}.

Recently the Wealth Thermalization Hypothesis (WTH) was proposed in
\cite{fpuarxiv2025} with the aim to explain the origins of
such universal wealth inequality existing in countries
and the whole world. The WTH argues that various types of links and nonlinear 
interactions between society members (agents) in a given country
leads to dynamical thermalization over layers of Social Stratification in
a Society (SSS).  This SSS is characterized by
wealth values $w_m$ of corresponding layer states with
approximately constant density of states $\nu= dm/dw_m =const$.
The appearance of SSS is broadly discussed in social science
starting from the Marxist theory of classes (see e.g. 
\cite{marx,lenski,sanders,kerbo,wikistratif}).
Using a mathematical SSS model it was shown in \cite{sssarxiv}
that links between agents $m$ and their nonlinear
interactions lead to dynamical chaos and Rayleigh-Jeans (RJ) 
thermalization.
In fact the wealth values $w_m$ can be considered as
energies $E_m$  of states $m$ (or related frequencies).
As argued in \cite{boghosian1,boghosian2} 
the thermalization process has two conserved quantities being total system
energy $E = \sum_m E_m \rho_m$ and probability norm 
$\eta = \sum_m \rho_m$ (number of agents)
where $\rho_m$ is the time averaged squared amplitude and occupation 
probability of each oscillator mode $m$ considered in \cite{sssarxiv}.  
According to the well known results
of thermodynamics of classical fields \cite{landau,zakharovbook}
the steady-state distribution of probabilities (or fractions)
for the RJ thermalization has the form:
\begin{equation}
\rho_m = \frac{T}{E_m-\mu} \; ({\rm RJ}) .
\label{eqrj}
\end{equation}
Here $\rho_m$ represent averaged stationary 
probabilities at certain wealth states 
 $0 \leq m < N$ with wealth/energies $w_m=E_m$. The 
two parameters correspond to the system temperature $T$ and 
the chemical potential $\mu(T)$ which are 
determined by the values of total energy $E$ and norm $\eta=1$
via both equations given above. (See e.g. Refs. 
\cite{fpuarxiv2025,sssarxiv,rmtprl} and below for more details.)

The RJ distribution also follows from the Bose-Einstein thermal
distribution of quantum bosonic fields
at high temperatures $T \gg E_m - \mu$ \cite{landau}:
\begin{equation}
 \rho_m=\frac{1}{\exp[(E_m-\mu)/T]-1} \; ({\rm BE}) .
\label{eqbe}
\end{equation}

It was shown that the dynamical RJ thermalization
appears in such systems as SSS models \cite{sssarxiv},
nonlinear perturbation of Random Matrix Theory (RMT) \cite{rmtprl}
and social networks \cite{fpuarxiv2025}, light propagation in multimode
optical fibers \cite{ourfiber}. This thermalization results from chaotic
dynamics emerging when a nonlinearity parameter exceeds a certain chaos border
\cite{chirikov1979,lichtenberg}. Below the chaos border the dynamics
is integrable in the main part of the phase space in agreement
with the Kolmogorov-Arnold-Moser (KAM) theorem \cite{arnold,sinai}.

In fact, RJ thermalization has been studied experimentally and numerically
for light propagation in a nonlinear media of multimode optical fibers
\cite{wabnitz,picozzi1,chrisrep,picozzi2,babin,chris,picozzi3}.
However, the origins of RJ thermalization were
never clarified there as well as its absence at weak nonlinearity
below the chaos border when the dynamics becomes
integrable as it is stated by the KAM theorem.
At the same time analytical studies
and numerical simulations of the Nonlinear
Sch\"odinger Equation (NSE) for light propagation \cite{picozzi1}
showed that at moderate nonlinearities and low energy
injected in the system the RJ thermal distribution (\ref{eqrj})
leads to an enormous condensation of total
probability at the fiber ground state 
with about 80-90\% of probability at this state. This phenomenon, called
a self-cleaning, was experimentally demonstrated
for fibers in \cite{wabnitz,picozzi2}.
The RJ condensation was also found in the NSE numerical modeling of 
light propagation in D-shape quantum chaos fibers \cite{ourfiber}.

Actually the RJ condensation follows directly 
from the RJ thermal distribution (\ref{eqrj})
and two integrals of motion being energy  $E$ and norm $\eta$
that can be considered as two constraints. Then it is easy to show that
at small temperature $T \sim E_g$ a huge macroscopic probability $\rho_g \rightarrow 1$
is condensed at the ground state $m=0$ with energy $E_g=E_0$.
Examples of such condensation in the RJ standard (RJS) model
with equidistant energy states $E_m = \nu m$ ($0 \leq m \leq N-1)$)
and constant density of states $\nu = dm/dE_m =const$
are given in \cite{fpuarxiv2025,ourfiber}. According to the WTH
the wealth inequality of households
in countries results from this RJ thermal condensation
when the total wealth (energy) of a country is small
compared to the wealth dispersion $B$ in the society,
where $B =E_{max}  \approx N/\nu$ is determined by a maximal wealth
of households in a country.

We note that the first results for RJ condensation at low energy modes
were obtained by H.~ Fr\"ohlich
in 1968 from (\ref{eqrj}) for electric modes in biocells \cite{frohlich1,frohlich2}.
However, his analysis contained a logical gap
since it was motivated by the quantum BE distribution (\ref{eqbe})
making a transition to the RJ case (\ref{eqrj}) assuming a high temperature
regime with $T \gg E_0, E_1$ getting RJ condensation in (\ref{eqrj})
which however exists only at $T \sim E_0, E_1$. 
Also in \cite{frohlich1,frohlich2}, it is assumed that
in the system there are external energy pumping and absorption 
so that the relation to the case of the conservative
RJ thermalization is not so direct (see discussion in \cite{ourfiber}).
More recently the emergence of RJ condensation from (\ref{eqrj})
was established in 2005 for nonlinear waves in NSE \cite{picozzi1}.
In these works, see e.g. \cite{picozzi1,chrisrep}, the origins of 
RJ thermal distribution were attributed to the Kolmogorov-Zakharov (KZ) 
turbulence \cite{zakharovbook,nazarenko}. 

The RJ condensation was experimentally observed in
multimode optical fibers in \cite{wabnitz,picozzi2}.
For the NSE in a quantum chaos fiber
it was shown \cite{ourfiber} that the  RJ condensation
takes place in a pure Hamiltonian dynamical system
due to dynamical chaos above the  chaos border
while below this border there is no thermalization
and dynamics is integrable
in agreement with the KAM theorem. Thus energy pumping and absorption,
essential for the KZ turbulence, are now required
and the RJ condensation appears in a purely Hamiltonian system.

De facto the RJ condensation is a specific case of
a more generic phenomenon known in statistical mechanics as
constraint-driven condensation \cite{trizac,satya,marsili}.
This condensation is universal and
exists for continuous systems such as
coalescence in granular media, jamming in traffic, 
gelation in networks \cite{satya}
and financial data analysis \cite{marsili}.
Due to this universality 
we argue that the RJ condensation is at the origin of
wealth inequality in countries, the whole world
and other society organizations.
We illustrate this universality
showing that the RJ thermalization and condensation
give a good description of real Lorenz curves
not only for wealth of households in countries but
also for the Gross Domestic Product (GDP) of countries;
market capitalization of companies at stock exchange in Hong Kong, London,
New York, Shanghai; bitcoin distribution between wallets (users);
world trade between countries. The appearance of dynamical RJ thermalization
due to linear links and nonlinear interactions
between agents have been demonstrated for the SSS models \cite{sssarxiv},
nonlinear perturbation of RMT \cite{rmtprl},
social networks \cite{fpuarxiv2025} and NSE evolution
in quantum chaos fibers \cite{ourfiber}.
Due to that we use the RJ expression (\ref{eqrj})
as given and compare the Lorenz curves from
this thermal distribution with the real Lorenz curves
for the wealth inequality in the world.
We also discuss the properties of the Pareto distribution
at very high wealth values.

The article is composed as follows:
Section 2 reminds the properties of the RJ thermal distribution and 
the construction of associated Lorenz and Pareto curves, 
the WTH description of Lorenz curves of households of countries and the world
is given in Section 3, the GDP Lorenz curves are analyzed in Section 4,
the WTH description of data of stock exchange markets
is presented in Section 5,
the bitcoin transactions are studied in Section 6,
the analysis of the world trade data from
UN COMTRADE is presented in Section 7,
the universality of the thermodynamic description
is highlighted in Section 8
and the conclusion is given in Section 9.
Additional data and detailed analytical results for two specific 
RJ models in the continuous limit $N\to\infty$ are given in the Appendix.

 \section{Properties of RJ thermal distribution and condensation,
   implications for Lorenz and Pareto curves}

\subsection{Features of RJ distribution} 

In this work, we consider two specific model spectra for the 
linear oscillator modes $E_m$. The first one is the RJS model with 
equidistant modes 
$E_m=m/N$, $m=0,1,\ldots, N-1$ with constant density of states 
$\nu_{\rm RJS}=dm/dE_m=N$ and the second one is called RJ Extended (RJE) 
model with a parameter $a$ and $E_m=(e^{am/N}-1)/(e^a-1)$. 
Both have a bandwidth $B=E_{N-1}-E_0\approx 1$ (note that in former 
work \cite{fpuarxiv2025,sssarxiv} we used the RJE model with a 
different bandwidth $B>1$ which is does not affect the construction of 
Lorenz curves; see below for details). 
The RJE model appears rather natural since its 
density of states decreases (for $a>0$) 
at high energies (wealth) values $E_m$ according to:
\begin{align}
\label{eqnu}
\nu_{\rm RJE}(E_m) = \frac{dm}{dE_m} = \frac{N(e^a-1)}{a(1+(e^a-1)E_m)}
\end{align}
and we consider that it 
is more realistic when we want to describe real data sets. 

In the limit $a \rightarrow 0$, we recover the RJS model
while with increasing $a$ the exponential growth of $E_m$  becomes
more dominant. We will see that the RJE model
gives typically a better description of real Lorenz and Pareto curves. 

First, we remind the basic RJ thermalization properties for 
generic spectra with a few figures for the RJS model as illustration. 
Let us assume that we have 
$N$ linear classical oscillators with individual energies 
$E_m$, $m=0,\ldots,N-1$ (e.g. spectrum according to the RJS or RJE model) 
which are coupled by some small 
non-linear perturbation (see 
Refs.~\cite{fpuarxiv2025,sssarxiv,rmtprl,ourfiber} for examples) 
such that there are two conserved quantities being the global (squared) 
amplitude and total energy:
\begin{align}
\label{eqS1}
\eta = \sum_{m=0}^{N-1} \rho_m =1 \quad,\quad E=\sum_{m=0}^{N-1} E_m\rho_m
\end{align}
where $\rho_m$ is the time averaged squared amplitude and occupation 
probability of each oscillator. If the nonlinear terms are 
weak or moderate (but above a certain chaos border)  or if there is some 
weak coupling to an external 
system (which respects both constraints (\ref{eqS1})) one can assume 
that the system is thermalized. 
Applying the framework of the grand canonical ensemble one introduces 
two parameters: temperature $T$ and chemical potential 
$\mu$ to satisfy both constraints (\ref{eqS1}) in average and it can be shown 
(see e.g. Ref.~\cite{rmtprl}) that 
\begin{align}
\label{eqS2}
\rho_m=\frac{T}{E_m-\mu}\quad,\quad T=\frac{E-\mu}{N}
\end{align}
where the expression for the temperature $T$ is obtained 
from $\sum_m (E_m-\mu)\rho_m=(E-\mu)$ which follows directly 
from (\ref{eqS1}). The chemical potential is determined 
(using standard numerical techniques) by solving the implicit
equation:
\begin{align}
\label{eqS3}
1=\frac{E-\mu}{N}\sum_{m=0}^{N-1}\frac{1}{E_m-\mu}
\end{align}
which allows for one physical solution of $\mu$ outside the 
energy interval $[E_{\rm min},E_{\rm max}]$ with either 
$\mu<E_{\rm min}$ ($T>0$) or $\mu>E_{\rm max}$ ($T<0$) 
(depending on the value of $E$ we have either $T<0$ or $T>0$) 
such that $\rho_m>0$. Without going into the details, we mention 
that for the limiting case $|\mu|\to \infty$, corresponding to $|T|\to\infty$, 
one can work out a simplified explicit approximate expression of $\mu$ 
(as a function of $E$ and $E_m$) showing that the transition 
from $T>0$ ($\mu<E_{\rm min}$) 
to $T<0$ ($\mu>E_{\rm max}$) appears exactly at the critical 
energy $E=E_C=(\sum_m E_m)/N$ (``center of mass'' of the spectrum $E_m$). 
In this work, we will only focus on the case $T>0$ with $E<E_c$ which is 
the most important case to fit the real data (e.g. of the World Lorenz curve). 

The data presented in this work were obtained by this procedure for two 
different model spectra and certain values of $N=10000$. 
Concerning the RJS model, we  also 
considered the cases $N=100$, $N=1000$ and verified that the obtained 
Lorenz curves are very close (in graphical precision). In the 
Appendix section A1 (A2), we 
also present an analytic theory of the RJS (RJE) model in the continuous limit 
$N\to\infty$ with explicit formulas for key quantities (e.g. Lorenz curves, 
Gini coefficient etc.; see below for details) in terms of the 
chemical potential which is determined by a specific implicit equation. 
The results for $N\to\infty$ are very close to the case of finite $N=10^4$. 

\begin{figure}[htbp]
\begin{center}
\includegraphics[width=0.8\textwidth]{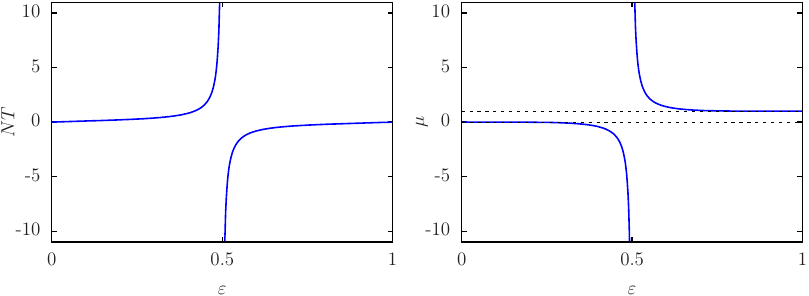}
\caption{\label{fig1}
The left (right) panel shows the (rescaled) temperature $NT$ 
(the chemical potential 
$\mu$) versus the rescaled energy 
$\varepsilon=E/B$ for the RJS model $E_m=m/N$, $N=10000$. 
The dashed black lines in the right panel correspond to the values of 
$E_0=0$ and $B\approx 1$ showing that either $\mu<E_0$ (for $T>0$) 
or $\mu>B$ (for $T<0$). 
}
\end{center}
\end{figure}  

Fig.~\ref{fig1} shows for the RJS model the dependence of $T$ and $\mu$ 
on $\eps=E/B\approx E$. 
Note that the left panel shows the rescaled temperature $NT$ since 
typical numerical values of $T$ are $\sim 1/N$ due to the finite 
value of $B$ in our particular model. 
The figure illustrates that $-\mu\to 0$ ($\mu\to -\infty$) for $\eps\to 0$ 
($\eps\to \eps_C=1/2$).

\begin{figure}[htbp]
\begin{center}
\includegraphics[width=0.65\textwidth]{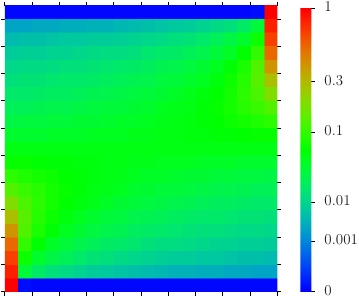}
\caption{\label{fig2}
Color plot of the coarse-grained thermalized occupation 
probabilities $\rho_m=T/(E_m-\mu)=(E-\mu)/[N(E_m-\mu)]$
for the RJS model. The $x$-axis corresponds 
to the fraction $E_m/B\in[0,1]$ (left to right) 
and the $y$-axis to the rescaled 
energy $\varepsilon$ (top to bottom 
for increasing values). The tics indicate integer multiples of 0.1 
for both quantities. 
The color values shown in the color bar 
correspond to the value of $\rho_m$ averaged over intervals of size 
$1/20$ . 
To increase visibility of small values a non-linear color bar scale 
is used (e.g. green color corresponds to $1/16$). 
}
\end{center}
\end{figure}  

\begin{figure}[htbp]
\begin{center}
\includegraphics[width=0.65\textwidth]{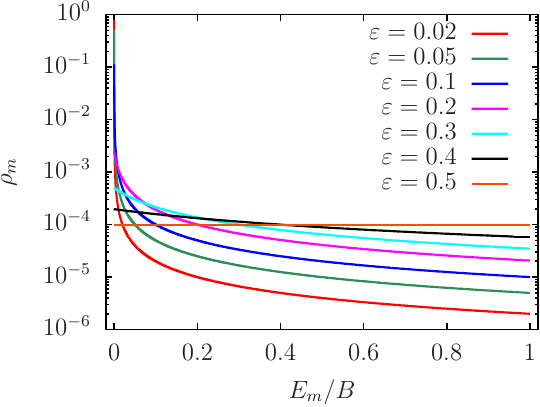}
\caption{\label{fig3}
Dependence of the thermalized occupation 
probabilities $\rho_m=T/(E_m-\mu)=(E-\mu)/[N(E_m-\mu)]$ on 
$E_m/B$ for the RJS model 
$E_m=m/N$, $N=10000$ and different values of $\eps$. 
}
\end{center}
\end{figure}

Using (\ref{eqS3}) one can show (for finite $N$) that 
$-\mu\approx E/(N-1)\ll E$ for very small energies $0<E\ll 1/N$ 
and in this particular case we have $\rho_0\approx 1$ (strong 
condensation) and other 
$\rho_m\sim E/(NE_m)\ll 1/N$ (for $m>0$). 
With increasing values of $E$ (or $\eps$) 
the values of ``$-\mu$'' increase and more probability is shifted to 
the other $\rho_m$ values for $m>0$. At $\eps\approx 1/2$ 
we have very large values of ``$-\mu$'' (and of $T$) such that all 
$\rho_m\approx 1/N$ are uniformly constant. Further increase of 
$\eps$ enters the regime of negative temperatures (with $\mu>E_{\rm max}$) 
with possible 
condensation at the last oscillator with $\rho_{N-1}\gg 1/N$ 
(in this work we do not consider the regime of $T<0$).
These features are visible in both figures 
Figs.~\ref{fig2} and \ref{fig3} showing 
$\rho_m$ versus $E_m/B$ for different values of $\eps$ (as color plot 
or curves in logarithmic scale). The effect of 
condensation for small $\eps$ with a finite macroscopic 
probability $\rho_0\gg 1/N$ is clearly 
visible in both figures and qualitatively one could even say that it extends 
even up to $\eps=0.2$ with $\rho_0=0.002495$ still being larger than $1/N$. 
However, there are also some other values of $\rho_m$ with small $m$ that are
significantly larger than $1/N$ (as can be seen in \ref{fig3}
for the first 5\% of modes with $\rho_m\ge 3/N$). Also the coarse-grained 
average value at the first 5\% of modes at $\eps=0.2$ is roughly 
0.35 times the 
maximal coarse-grained value at $\eps\approx 0$ (according to 
Fig.~\ref{fig2}). This effect corresponds to (modest) condensation 
on several modes or on a given small mode interval.

\begin{figure}[htbp]
\begin{center}
  \includegraphics[width=0.9\textwidth]{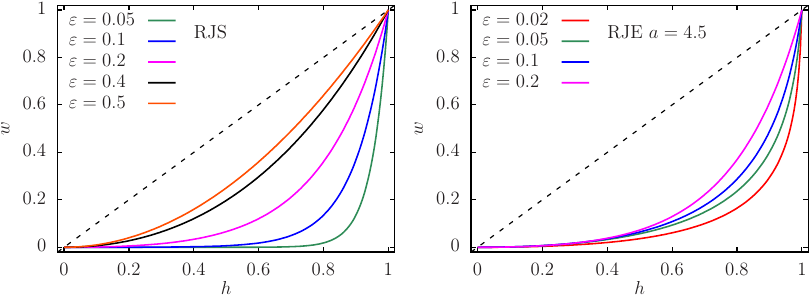}
\caption{\label{fig4}
Left panel: Lorenz curves for the RJS model 
for $\eps=0.05,0.1,0.2,0.4,0.5$ (bottom to top). 
The $x$-axis corresponds to the 
cumulated fraction of households ($h$) and the $y$-axis to
the cumulated fraction of wealth ($w$). 
The dashed line is the line of perfect equipartition $w=h$. 
The respective values of the Gini coefficients $G$ for all curves are 
$G=0.900,0.800,0.625,0.407,0.333$. 
Right panel: Lorenz curves for the RJE model at $a=4.5$ 
for $\eps=0.02,0.05,0.1,0.2$ (bottom to top) with respective 
Gini coefficients $G=0.798,0.725,0.678,0.615$.
Note that for these two cases
the critical value of $\eps$ for the transition of positive temperature 
(with $\mu<0$) to negative temperature (with $\mu>B=1$) is at $\eps_C=0.5$ 
(left panel) and $\eps_C=0.211$ (right panel) and that the RJS Lorenz 
curve at $\eps=0.5$ is given by the simple expression $w=h^2$. 
In this figure, we use for both RJS and RJE models 
the continuous limit $N\to\infty$ but the curves for finite $N=10000$ 
are identical on graphical precision (differences typically below $10^{-4}$; 
see also Appendix Figures~\ref{figA1} and \ref{figA2}). 
We remind that the RJS model is a special case of the RJE model at $a=0$. 
}
\end{center}
\end{figure}

\subsection{Construction of the Lorenz curve for RJ models}

To construct for a generic RJ model, such as the RJS or RJE model, 
the Lorenz curve, we need to define the cumulated household 
variable $h$ and the cumulated wealth variable $w$ 
(both with maximal value normalized to unity). For this, we assume 
that the probabilities $\rho_k$ represent a given fraction of households 
or population associated to the mode $k$. Therefore, for the cumulated 
household variable, 
we add the probabilities $\rho_k$ up to some index value $m$ and for the 
cumulated wealth variable we add $(w_k/w_s)\rho_k$ in a similar way, 
where $w_k=E_k$ (wealth/energy value of agent/mode $k$) and 
$w_s=E=\sum_{l=0}^{N-1} w_l\rho_l$ (average wealth/energy over 
all agents/modes with probabilities $\rho_l$). 
More explicitely, we construct the Lorenz curve 
as the set of points $(h(m),w(m))$ for $m=0,1,\ldots, N$ with 
the partial sums 
$w(m) =  \sum^{m-1}_{k=0} w_k \rho_k/w_s$ and 
$h(m) = \sum^{m-1}_{k=0} \rho_k$ such that $0 \leq h,w \leq 1$. 
Here the maximal value of $h$ and $w$ is $w(N)=h(N)=1$ 
since $w_s=\sum_m w_m\rho_m$ and $\sum_m \rho_m=1$. 
(Note that this procedures gives $N+1$ data points for a spectrum 
with $N$ modes due to the trivial initial value $h(0)=w(0)=0$). 
This procedure can be implemented directly for a given spectrum 
at finite $N$, provided we have first determined (for the 
given energy $E$) the 
chemical potential $\mu$ which fixes $\rho_k$ by (\ref{eqS2}). In both 
Appendix sections A1 and A2, we compute these partial sums analytically 
in the limit $N\to\infty$ as integrals resulting in explicit expressions 
for the Lorenz curves $w(h)$ which are 
Eq. (\ref{eqlorA}) for the RJS model and 
Eqs. (\ref{eqsh2}), (\ref{eqRJEwofh}) for the RJE model. However, these 
expressions still depend on the chemical potential which is to be determined 
by the suitable implicit equation (for given values of $E$ and also 
$a$ for the RJE model; see Appendix A1 and A2 for more details). 

Examples of Lorenz curves for the RJS and the RJE model (at $a=4.5$) 
at various values of the rescaled average energy $\varepsilon =E/B\approx E$
(here $B\approx 1$) are shown in Fig.~\ref{fig4} (left or right panel 
for RJS or RJE with $a=4.5$ respectively). Here, we show 
the Lorenz curves obtained by the analytical expressions of the appendix 
for the limit $N\to\infty$ but we have verified that the curves for 
finite $N=10^4$ are identical on graphical precision with an error 
similar or less than $10^{-4}$ which is illustrated in Appendix 
Figs.~\ref{figA1} and \ref{figA2}. 

For both models, the curves of Fig.~\ref{fig4} show that at smaller 
values of $\eps$ (lower values of $T$) we have larger values of 
the Gini coefficient $G$ closer to unity (stronger inequality). 
For the RJS model 
at largest possible value of $\eps=\eps_C=0.5$ (or more precisely slightly 
below $\eps_C$ to have a stable solution for $\mu$), 
we have the simple expression $w=h^2$ with $G=1/3$ which 
can be understood by the fact that this case corresponds to $T\to\infty$ 
with $\rho_l=1/N=$const. (see also the orange line at $\eps=0.5$ 
in Fig.~\ref{fig3}) such that $h(m)=m/N$ and 
$w(m)=\sum_{l=0}^{m-1} l/(\eps_C N^2]\approx (m/N)^2$ 
(see also Eq. (\ref{eqlor2}) and subsequent discussion in Appendix A1). 

For the continuous RJS model and small $\eps\le 0.2$, we have (for 
$N\to\infty$) also the very simple 
quite accurate approximations $w(h)\approx e^{(h-1)/\eps}$ and 
$G\approx 1-2\eps$ combined with $\mu\approx -e^{-1/\eps}\ll \eps$ 
(see Appendix A1 and in particular 
Appendix Fig.~\ref{figA1} for more details on this).

The parameter choice $a=4.5$ for the RJE Lorenz curves in the right 
panel of Fig.~\ref{fig4} corresponds to a typical value which gives 
a good fit with real data (see next section). For this case, the critical 
energy $\eps_C=1/a-1/(e^a-1)\approx 0.211$ for the transition to 
negative temperatures is significantly lower as for the RJS case since 
here there are more levels at smaller energies resulting in a reduced 
energy center of mass. 

To compare the results of the RJS and the RJE model with real Lorenz curves, 
we apply the following procedure. First, for both cases we
fix the rescaled energy $\varepsilon = E/B$ by the condition
that the Gini coefficient $G$ for the RJS/RJE model 
is equal to its value of the real Lorenz curve. This is a simple and 
efficient choice which is possible since $G$ is a decreasing function 
with respect to increasing values of $\eps$ as can be seen in both panels 
of Fig.~\ref{fig4}. 
Concerning the parameter $a$ of the RJE model, we minimize a certain 
functional that measures the geometrical (orthogonal) curve distance between 
both curves (under the constraint that for each value of $a$ 
the value of $\eps$ is recomputed by the condition of identical Gini 
coefficients between both curves). We do not enter into more details of 
this procedure which is more complicated than a simple fit and 
which requires three levels 
of iterations to determine $\mu$, $\eps$ and $a$. 
We have implemented this procedure for both cases of 
finite $N$ and the continuous limit $N\to\infty$ with very similar results. 

We have also observed that modest variations of $a$, e.g. from $a=4.5$ to 
$a=4.4$, provide only very minimal modifications of the resulting Lorenz 
curves, provided that the value of $\eps$ is recalculated to match 
the initial Gini coefficient. We have also tested different variants of this 
procedure where the curve distance is minimized in logarithmic scale or 
for the Pareto curve in double logarithmic scale 
(see next subsection) which provide 
sometimes quite different solutions for the parameter $a$. However, 
for simplicity, we do not enter into more details on this and all results 
for optimal $a$ values for the RJE model with respect to some real data 
presented in this work were obtained for the Lorenz curve in normal scale. 

\subsection{RJ implications for the Pareto curve}

The Lorenz curve describes the global wealth distribution and 
it focuses mostly on the main bulk of the whole population, 
e.g. for the world population \cite{piketty1,piketty2} the typical 
value $w(0.5)\approx 0.02$ shows that about 50\% of the lower part 
of the population owns only 2\% of the whole wealth. 
However, in economy studies of inequality there is also
a significant interest in the properties of the small oligarchic 
fraction of population 
(usually about a few percent) that owns a huge amount
of the total wealth (about 50\%; see e.g. \cite{zucman1}).
A special analysis of this small part of the Lorenz curve
is usually described by the Pareto curve  \cite{pareto}
which assumes a power law distribution decay
of owners with high wealth (see e.g. \cite{yakovenko2}).

For practical analysis, it is therefore also useful to consider the 
cumulative distribution function (CDF) $C(w_m)$ which gives the 
fraction of households (companies, people, countries etc.) having 
a wealth larger than $w_m$ and to show it in a double logarithmic 
representation such that eventual power law tails are well visible. 
Mathematically, the case of the expression Pareto distribution corresponds to 
the special form $C(w_m)=1$ for $w_m>w_0$ and $C(w_m)=(w_0/w_m)^{-\alpha}$ 
with two parameters $w_0>0$ and (typically) $\alpha>1$. In our work, 
we will simply use the notation 
{\em Pareto distribution}  for the CDF $C(w_m)$, 
for other more general cases where there is not (necessarily) a 
simple power law, e.g. the cases of the RJS and RJE where it is possible 
to compute this function either numerically for finite $N$ or analytically 
in the limit $N\to\infty$. 

We mention that one can also study the usual probability density function 
(PDF) in $w_m$ defined by $p(w_m)=-C'(w_m)$ such that 
$p(w_m)\,dw_m=C(w_m)-C(w_m+dw_m)$ is the fraction 
of households with wealth in the interval $[w_m,w_m+dw_m]$. 
However, in this work we will not use the PDF and focus more 
on the CDF being called the Pareto curve (together with the Lorenz curve). 

From $C(w_m)$, one can compute the 
Lorenz curve $w(h)$ in the following way: first one defines 
the cumulated household quantity $h(w_m)=1-C(w_m)$ (with $0\le h\le 1$) 
which corresponds to the fraction of (poorest) households having each 
an individual household wealth less than $w_m$. 
Then one expresses $w_m=w_m(h)$ as a 
function of $h$ and computes the integrated (and rescaled) 
cumulated wealth by $w(h)=W(h)/W(1)$ with 
$W(h)=\int_0^h w_m(\tilde h)\,d\tilde h$ such that $w(0)=0$ and $w(1)=1$. 

For example, for the exponential Boltzmann-Gibbs (BG) distribution 
at temperature $T>0$ one has: 
\begin{align}\nonumber
C_{\rm BG}(w_m)&=e^{-w_m/T}\folgt
h=1-e^{-w_m/T}\folgt w_m(h)=-T\ln(1-h)\folgt\\
\nonumber
W(h)&=-T\int_0^h \ln(1-\tilde h)\,d\tilde h = 
T[(1-h)\ln(1-h)+h]\end{align}
which gives the Lorenz curve 
\begin{align}\label{lorBG}
w_{\rm BG}(h)&=(1-h)\ln(1-h)+h\ .
\end{align}
This result was previously obtained in Ref. \cite{yakovenko2}. 
It respects the limiting cases $w(0)=0$ and $w(h)\to 1$ 
for $h\to 1$ and it no longer depends on $T$, i.e. the BG-Lorenz curve 
has no parameter. In particular, it has a fixed value for the 
Gini coefficient which 
can be directly computed from (\ref{lorBG}) as $G=1-2\int_0^1 w(h)dh=\frac12$.
Such a value is rather high
(in comparison to typical real data) with a low degree 
of inequality. On the other hand, the derivative $w'(h)$ diverges 
(logarithmically) for $h\to 1$ which is a manifestation that this 
model does not have a finite limit for $w_m$ which can take arbitrarily large 
values but with exponentially small probabilities such that the cumulated 
wealth stays finite. For real data, the RJS and RJE models we have a finite 
limit $w_m\le w_{\max}$ such $w'(1)$ may be large but is still finite (we 
typically apply a rescaling of units such that $w_{\max}=1$). 
Furthermore, in real data, we have typically larger Gini coefficients 
$G\sim 0.7-0.9$ or even $G\approx 0.95$ for some cases of 
Stock market or Bitcoin (see below). 
Globally, the Lorenz curve and $C_{\rm BG}(w)$ for the BG model fit very 
badly real data (see next section for two examples).
In \cite{yakovenko2}, the fact that the BG distribution always has
$G=0.5$, that is very different from real situations, 
was masked by the manual introduction of a complementary
(power law) Pareto distribution at high wealth that gave
a variation of $G$ depending on a subjective fraction of the Pareto tail.

We also point out 
another important difference between the BG description \cite{yakovenko2}
and RJ model. Thus for a gas of atoms the BG 
distribution $\rho_m \sim \exp(-E_m/T)$
describes fluctuations of kinetic energy of atoms  $E_m$ while the average
steady-state energy of each atom is the same being $\langle E_m\rangle=3T/2$ 
(in a 3D geometry). 
In contrast to this the RJ distribution (\ref{eqrj}) gives already 
steady-state thermal averaged probabilities for given modes 
$m$ that are more stable. 

For the RJ distribution (\ref{eqrj}), the construction of 
the Pareto curve (at finite N) is straightforward by the identification 
$C(w_m)=\sum_{k=m}^{N-1} \rho_k$ (corresponding to $1-h(m)$) 
and $w_m=E_m/E_{N-1}$ if we fix $w_{\max}=1$. 
In Appendix subsection A2.6, we also compute the Pareto distribution 
analytically for the RJE model in the continuous limit $N\to\infty$ 
which gives the precise expression (\ref{eqCw1}) together with 
the RJS limit (\ref{eqCw0}) obtained for $a\to 0$. Here, we only mention 
the nice simplified form:
\begin{align}
\label{eqCw2main}
C_{\rm RJE}(w_m)&\approx
\frac{\eps-\mu}{a}\left(\frac{1}{w_m}-1\right)
\end{align}
which is typically valid for rather large values of $a$ (e.g. $a>7$), 
small values of $\eps$ (with $G\sim 0.8$-$0.95$) and the 
interval $w_m\in[w_{\min},1]$ where $w_{\min}\ll 1$ 
is some small value (see the 
Appendix for more details). If we exclude from this interval values 
close to unity, i.e. for $w_{\min}<w_m\ll 1$, the above expression becomes 
even a simple power law $C_{\rm RJE}(w_m)\approx C_2 w^{-1}$ 
with exponent $-1$ and constant $C_2=(\eps-\mu)/a$. However, we will see that 
this simple power law is only visible for the cases with quite large 
values of $a$ and small values of $\eps$ where we have a significant 
interval for its validity. 

\section{WTH for Lorenz curves of countries and the world}

In this Section, we compare the WTH theory, based on the RJS and RJE models,
with the real Lorenz and Pareto curves of wealth inequality for households
of countries and the whole world. 

\begin{figure}[htbp]
\begin{center}
  \includegraphics[width=0.9\textwidth]{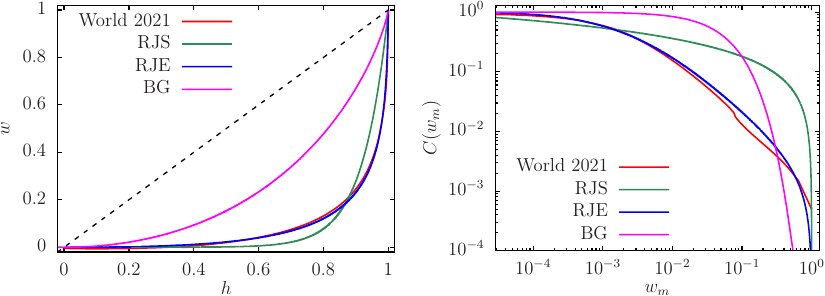}
\caption{\label{fig5}
Left panel: Lorenz curve of
cumulated wealth distribution $w$ vs cumulated fraction of households $h$
for the whole world in 2021  with Gini coefficient
$G=0.843$ (red curve with data taken from \cite{piketty2}), 
for the RJS model at $\eps=0.0784$ (green), for the RJE model at 
$a=4.74, \eps=0.0113$ (blue) and for the BG model (pink) using the expression 
(\ref{lorBG}) which is obtained in \cite{yakovenko2} and 
does not depend on $T$. The dashed black line shows the line 
of perfect equipartition $w=h$. 
Right panel: Pareto probability $C(w_m)$ versus individual rescaled wealth $w_m$ 
for the same cases and parameters as in the left panel. 
Here for the BG model $C_{\rm BG}(w_m)=e^{-w_m/T}$ depends on 
$T$ and the shown pink curve corresponds to 
$T = 0.0574$ obtained by the fit of the function $-w_m/T$ with 
$\ln(C(w_m))$ for the World data. 
The values of $\eps$ for the RJS and RJE models are obtained from the 
condition 
that the corresponding Lorenz curves of these models have the same 
Gini coefficient $G=0.843$ as the World data. For the RJE model 
the parameter $a$ is determined by minimizing the geometric curve distance 
from the World Lorenz curve at fixed $G$. 
Here for both RJS and RJE models we use
the continuous limit $N\to\infty$ but the curves for finite $N=10000$ 
are identical on graphical precision (differences typically below $10^{-4}$) 
and provide very close values for $\eps$ and $a$. 
}
\end{center}
\end{figure}

In Fig.~\ref{fig5} (left panel) we show the Lorenz curve for wealth of 
households of the whole world in 2021. This curve is obtained from the 
data presented in \cite{piketty2}
(from the front page of this web site with interpolated cell data
and best possible precision). 
For this data the Gini coefficient is rather high being $G=0.843$. 
The Lorenz curve from the RJS model with the same $G$ value clearly shows 
the presence of RJ condensation
corresponding to the phase of poor households 
when 50\% owns only 0.17\% of total wealth while
from \cite{piketty2} this fraction is 2\%.
However, even if the RJS model clearly captures the existence
of condensate of poor owners
its shape differs from the real Lorenz curve obtained from data 
of \cite{piketty2}.
Due to that we show also the Lorenz curve for the RJE model
with fit parameters $\varepsilon$ and $a$ which gives a very close curve 
to the real data of \cite{piketty2}. 
The left panel also shows the Lorenz curve from the 
BG theory \cite{yakovenko2}
which is very different from the real data. 

In the right panel of Fig.~\ref{fig5} we show the Pareto curve
$C(w_m)$ for the real Lorenz data \cite{piketty2}
of left panel. The data clearly shows that the BG theory does not work at all. 
Here the BG curve still depends on the BG temperature with a certain value 
$T=0.054$ used in Fig.~\ref{fig5} but other values (using different fit 
procedures) are not better. 

The RJS model describes only high values $0.4 < C(w_m) \leq 1$
corresponding to small values of cumulated wealth $w$ of poor households $h$
(see left panel) 
but it fails to describe the households at high wealth. 
Indeed, the RJS model captures the condensation
of households with low 
wealth but it is not well suited for a description of rich households.
However, the RJE model with the optimal parameter $a=4.74$ 
describes well the Pareto curve
including the oligarchic phase with very high wealth
up to the very low Pareto probability with $C(w_m) \approx 7 \times 10^{-4}$. 
We point out that the Pareto curve
shows a certain curvature and cannot be described by
a simple power law tail as it is usually
expected (see e.g. \cite{yakovenko2}). For the RJE model the 
simplified analytical expression (\ref{eqCw2main}) agrees with the 
precise blue curve (and quite well with the red data curve) for 
$w\ge w_{\min}\approx 0.04$. Here, we have theoretically a power law 
with exponent $-1$ for the very short interval $w\in[0.04,0.1]$ 
which is roughly the slope of the blue curve in this interval 
(in double logarithmic representation) 
but globally this interval is simply too small to claim of a power law 
behavior for this case. 

\begin{figure}[htbp]
\begin{center}
  \includegraphics[width=0.9\textwidth]{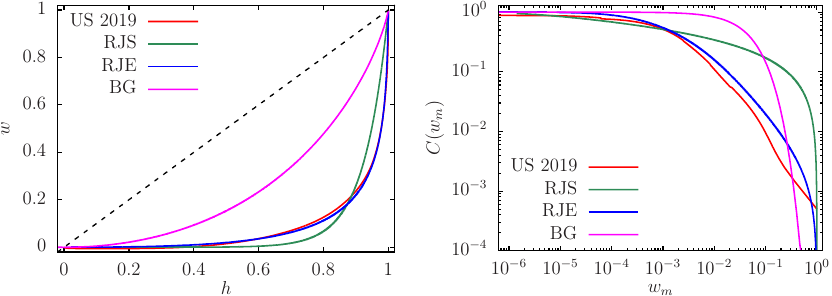}
\caption{\label{fig6}
Same as in Fig.~\ref{fig5} but for US 2019 with real data from 
\cite{usa2019}. The parameters are $G=0.852$, $\eps(RJS)=0.0742$; 
$\eps(RJE)=0.0105$, $a=4.72$, and for BG case in right panel $T = 0.0519$. 
}
\end{center}
\end{figure}

In Fig.~\ref{fig6} we also show  the Lorenz and
Pareto curves for USA in 2019 taken from \cite{usa2019}.
As for the case of Fig.~\ref{fig5} the BG theory does not work,
the RJS model captures the presence of condensation of poor households
but has rather visible deviations from
the real data for the Lorenz curve and even more for the Pareto curve.
The RJE model
describes well the Lorenz curve data and also the Pareto ones
for $C(w_m) > 0.01$. For $C(w_m) <0.01$ there are visible deviations
from real data even if the qualitative decay trend is captured. 
Here, the analytical expression (\ref{eqCw2main}) agrees with the 
RJE blue curve for $w\ge w_{\min}$ with the same value 
$w_{\min}\approx 0.04$ as for the World 2021 case which is plausible 
since the obtained fit values of $a$ and $\eps$ are indeed very 
close for both cases. 

We should note that the real data available at \cite{piketty2,usa2019}
does not have sufficiently fine cells so that
we used a smooth (rational) interpolation of data that
is not very accurate at high wealth $w_m$ values
that are especially important for the Pareto curve.
We discuss below the properties of Pareto curves
in more detail for the cases of stock exchange markets and
bitcoin transactions where the cell size is much
smaller.

\begin{figure}[htbp]
\begin{center}
\includegraphics[width=0.65\textwidth]{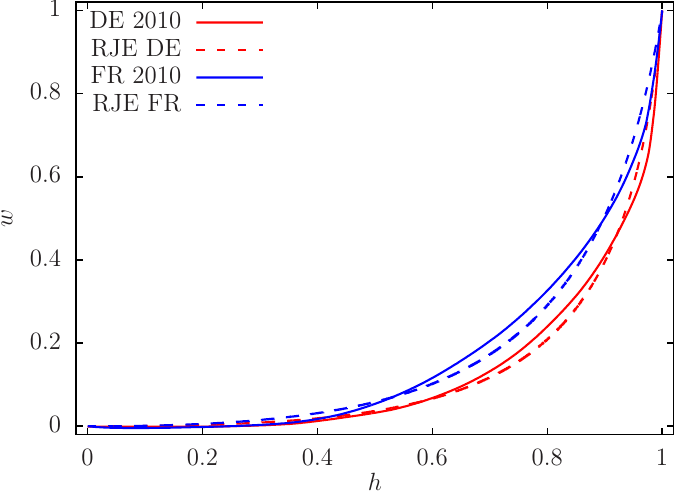}
\caption{\label{fig7}
Lorenz curves for wealth of DE (full red curve) and FR (full blue curve) 
in 2010 using data of \cite{defr}. The dashed lines of same color correspond 
to the Lorenz curves of the RJE model with optimal value of $a$ and 
same Gini coefficient as the reference data of either DE or FR. 
The parameters are $G=0.750,\, a=2.11,\,\eps=0.0682$ (DE) and 
$G=0.676,\,a=1.43,\,\eps=0.119$ (FR).  
}
\end{center}
\end{figure}

Finally, in Fig.~\ref{fig7} we present the comparison of real Lorenz curves,
taken from \cite{defr} for Germany and France in 2010, with the RJE model.
It shows reasonable agreement between real data and the RJE theory even if
there are certain deviations being $\approx 0.014$ (11\%) (DE) and 0.032 
(16\%) (FR) in values of $w(h)$ e.g at a given $h=0.7$. In 
these cases the Lorenz data
are obtained with best possible precision from data figures of \cite{defr} 
combined with a smooth interpolation between raw data points extracted 
from these figures. 

\section{WTH for Lorenz curves of GDP of countries}

Gross Domestic Product (GDP) is a monetary measure of
the total market value of all final goods and services produced by a country
during a year \cite{gdpwiki}. International organizations
as United Nations (UN), International Monetary Fund (IMF)
and World Bank (WB) yearly report GDP values for
world countries and territories as e.g in \cite{gdpcountries}
for recent years (with up to $N=212$ countries).
GDP country values for years 1970-2023
are publicly available from the UN statistics division at \cite{unyears}
and we mainly use this data here. If we consider countries as
independent players with equal rights according to the UN convention
then a striking feature of the GDP distribution of countries 
is a strong inequality
when 50\% of all countries own only 1\% of the total GDP,
while 10\% (1\%) of countries own 82\% (44\%) of the total GDP.
(e.g. in 2023 \cite{gdpcountries,unyears}).
It is interesting to note that this inequality of the GDP distribution is
very similar to the inequality of wealth distribution for households
in countries (see numbers given above and \cite{piketty1,piketty2,boston}).

\begin{figure}[htbp]
\begin{center}
  \includegraphics[width=0.65\textwidth]{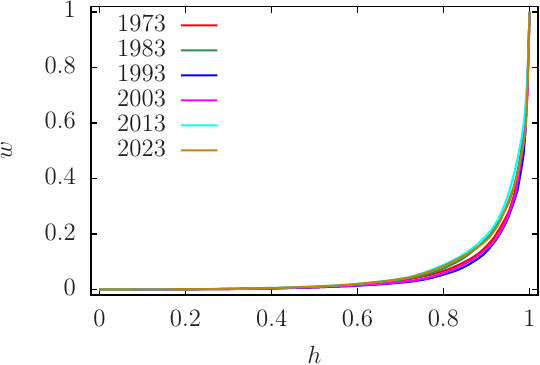}
\caption{\label{fig8}
Lorenz curves of GDP for countries from UN data \cite{unyears} 
for 6 years between 1973 and 2023. The $x$-axis corresponds the 
cumulated fraction of households/countries ($h$) and the $y$-axis to
the cumulated fraction of wealth/GDP ($w$). 
}
\end{center}
\end{figure}

The GDP Lorenz curves for years 1973, 1983, 1993, 2003, 2013, 2023,
obtained from UN data  \cite{unyears}, are shown in Fig.~\ref{fig8}. 
In this time range of 50 years
the Lorenz curve remains remarkably stable showing only small variations from year to year
even if the total world GDP is changed enormously from $5.23\times 10^{12}$  in 1973 to 
$1.05\times 10^{14}$ USD in 2023 
and the total number of countries and territories  is changed from $N=187$ to 
$N=212$ during this period
(the list of all countries is available at \cite{gdpcountries,unyears}).
The stability of the Lorenz curve confirms an adiabatic variation 
of the GDP world evolution and the approximate conservation of the 
two integrals of motion. 

\begin{figure}[htbp]
\begin{center}
  \includegraphics[width=0.9\textwidth]{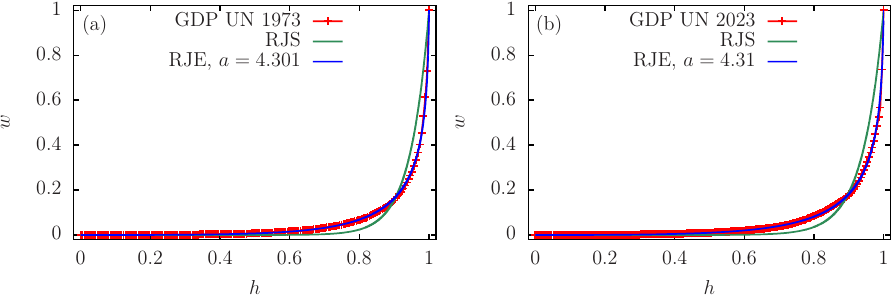}
\caption{\label{fig9}
  Panel (a): Lorenz curve for the year 1973 from UN data \cite{unyears}
  shown by red curve (with + symbols), the Gini coefficient is $G=0.892$;
  WTH theory with RJS model with $\varepsilon(RJS) = 0.0538$
  at same $G$ (green curve); RJE model results are shown with blue curve
  at $\varepsilon(RJE) = 0.00888$, $a=4.301$; panel (b) shows in the same 
  style, data for the year 2023 with $G=0.88$, $\varepsilon(RJS) = 0.0601$ 
  for the RJS model
and  $\varepsilon(RJE) = 0.01$, $a=4.31$ for the RJE model. Parameters
$T$, $\mu$ for RJS and RJE models for year 2023 at $N=212$
are given in text. 
}
\end{center}
\end{figure}

In Fig.~\ref{fig9}, we compare the GDP Lorenz curves of 1973 and 2023 with 
the WTH theory within the RJS and RJE models. The Lorenz curves for 1973
and 2023 years are rather close to each other
and thus the parameters of RJS and RJE models are also close
for these 2 years (same is valid for the other 4 years
shown in  Fig.~\ref{fig8}). On a scale of 50 years
the system parameters have only modest variations 
in the ranges:
$0.871 \leq G \leq 0.904$; $0.0479 \leq \varepsilon(RJS) \leq 0.0646$;
$0.0078 \leq \varepsilon(RJE) \leq 0.0157$; $3.622 \leq  a \leq 4.31$. 
The RJS model describes well the global 
behavior of Lorenz curves but has visible deviations while the RJE model 
gives almost perfect agreement with real data. 

Of course one can express a criticism saying that
a fit of a monotonic curve with two parameters (as for RJE curve)
may give a rather good agreement with the parameter $a$ 
influencing the form of the spectrum $E_m$ and 
the other parameter $\varepsilon$ giving  the 
rescaled energy. However, 
in our opinion the most important point is not the almost perfect
agreement of two curves of real data and the RJE model
(even if it is useful to have it)
but the physical origin of RJ condensation
that naturally explains the appearance of the 
phase of high poverty and the oligarchic phase. 
The physical reason of RJ condensation is a small rescaled value of total
system energy $\varepsilon = E/B \sim 0.01 \ll 1$.

We give the approximate values of system
parameters $B, T, \mu$ for the two cases 
of the RJS and the RJE model for the UN data of 2023 year in 
Fig.~\ref{fig9}(b). 
For this year we have $N=212$ countries and territories
with the total world GDP being $E \approx 100$ trillion USD $=10^{14}$ USD. 
For the subsequent discussion, we consider the RJS and RJE model at the same 
value $N^{\rm RJS/RJE}=212$ using the parameters values for $\eps$ and $a$ 
of Fig.~\ref{fig9}(b) (the curves in this figure correspond 
to $N^{\rm RJS/RJE}=10000$).
For the RJS model at $\varepsilon(RJS) =E/B =0.0601$ this 
gives $B =E/\varepsilon \approx 1670$ trillion USD
with a spacing between levels $\Delta(RJS) = E_{m+1} -E_m =B/N  \approx 7.9$ 
USD trillions.
The solution of the two equations for the two conserved integrals:
$\sum^{N-1}_{m=0} E_m \rho_m =E$ and
$\sum^{N-1}_{m=0} \rho_m=1$ with $\rho_m=T/(E_m-\mu)=(E-\mu)/[N(E_m-\mu)]$ 
gives $\mu = -0.74$ trillion USD, 
$T= (E-\mu)/N\approx E/N\approx 0.47$ trillion USD $\ll \Delta $
and the probability of the ground state 
is $\rho_0 = -T/\mu\approx 0.64$ indicating a rather strong RJ 
condensation. \\
For the RJE model we have 
$\mu=-1.57$ trillion USD, $T\approx E/N\approx 0.47$ trillion USD and 
$\rho_0=-T/\mu \approx 0.30$. Here the initial level spacing 
(at $m\ll N=212$ and using $a=4.31$) 
is $\Delta(RJE)=C/N=Ba/[N(e^a-1)]\approx B(RJE)/(17N)=E/(17 N \eps(RJE))
\approx (6/17)\Delta(RJS)\approx 2.8$ trillion USD which is still 
larger than $T$ but smaller then $\Delta(RJS)$.
Despite of the smaller value of $\varepsilon(RJE)=0.01$ the effect of 
RJ condensation for the RJE model in comparison to the RJS case is a 
bit reduced, but still clearly present since the effect of the reduced 
value of $\varepsilon$ (gives a factor $6$ for $\Delta$) is 
more than compensated by the the factor $1/17$ from the exponential 
factor $a/(e^a-1)\approx 1/17$ due to the RJE model. 

\begin{figure}[htbp]
\begin{center}
  \includegraphics[width=0.75\textwidth]{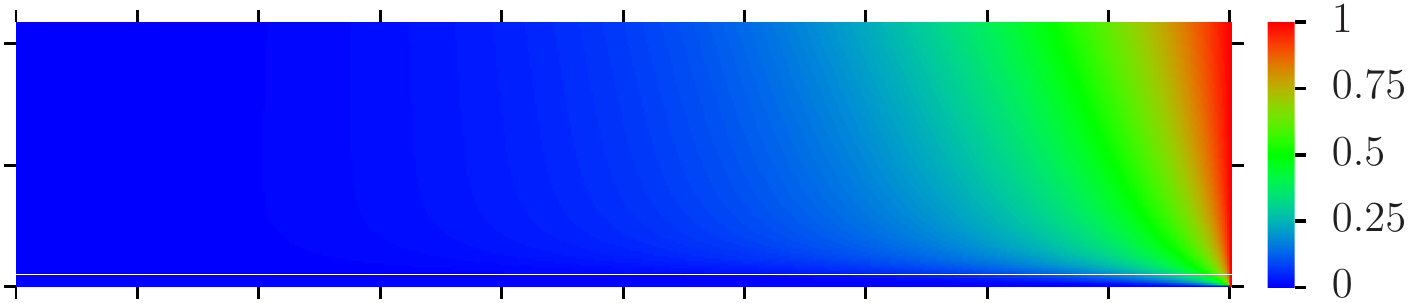}
\caption{\label{fig10}
Color plot of wealth $w$ from Lorenz curves of the RJE model 
at $a=4.31$. The $x$-axis corresponds to 
the fraction of households $h\in[0, 1]$ 
and the $y$-axis to the rescaled energy  
$\varepsilon\in[0, \varepsilon_C[$ 
where $\varepsilon_C=0.218$ is the critical value 
at which the transition from $T>0$ to $T<0$ appears. 
The ticks mark integer multiples of 0.1 
for $h$ and $\varepsilon$. The white line corresponds to 
$\varepsilon(RJE)=0.01$ obtained from the fit of the RJE model 
using the GDP data of 2023 shown in Fig.~\ref{fig9}. 
}
\end{center}
\end{figure}

It is possible to decrease the fraction of poverty phase by increasing
the dimensionless system energy $\varepsilon(RJE)$ as it is shown in 
Fig.~\ref{fig10} for the RJE model at year 2023
(at the same time the structure of energies $E_m$ is fixed with 
$a=4.31= const$ 
being independent of $\varepsilon(RJE)$). Indeed, the
results of Fig.~\ref{fig10} show that an increase of $\varepsilon(RJE)$
gives a significant reduction of the poverty phase (shown by blue color)
however this requires a quite strong boost of this $\varepsilon(RJE)$
parameter which may not be an easy task.
We note that here, as in \cite{fpuarxiv2025} for wealth of country households,
we consider only the cases with positive temperature
which for the RJE model at $a=4.31$ corresponds to 
 $\varepsilon(RJE) < \varepsilon_C=0.218$.

\begin{figure}[htbp]
\begin{center}
  \includegraphics[width=0.65\textwidth]{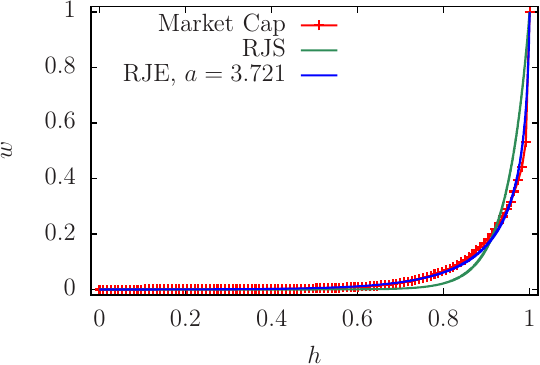}
\caption{\label{fig11}
  The Lorenz curve for SE market capitalization of $N=113$ countries,
  data from \cite{marketcap}. The Lorenz curves are shown
  in the same style as in Fig.~\ref{fig9}. Here $G=0.895$,
  $\varepsilon(RJS) =0.0525$ for the RJS model,
  $\varepsilon(RJE) =0.0119$, $a=3.721$ for the RJE model.
}
\end{center}
\end{figure}

\begin{figure}[htbp]
\begin{center}
  \includegraphics[width=0.75\textwidth]{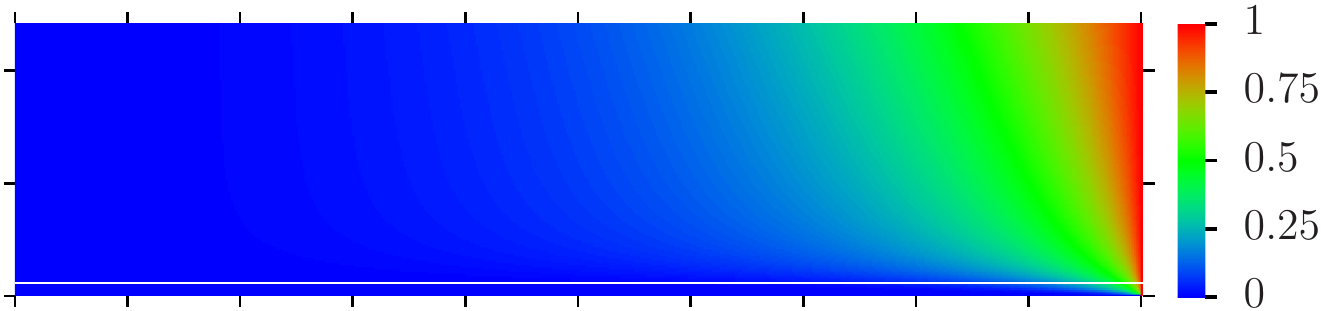}
\caption{\label{fig12}
As Fig.~\ref{fig10} with $a=3.721$, $\varepsilon_C=0.244$ and 
a white line at 
$\varepsilon(RJE)=0.0119$ obtained from the fit of the RJE model 
using the SE Market Cap data shown in Fig.~\ref{fig11}.
}
\end{center}
\end{figure}

Another wealth measure of a country can be expressed
via the total Stock Exchange (SE) market capitalization of all domestic companies 
listed in the country SE
(with World Bank data available at Wikipedia \cite{marketcap}).
In recent years 2024-2025 this list has $N=113$ countries
with the top total Market Cap $M=62186$ billion USD for USA
and minimal one $M=0.388$ million USD for Mongolia.
The ratio of Market Cap to GDP changes from country to country
being about 2.1 and 0.6 for USA and China, 0.8 for UK, 1.3 for France 
and 0.4 for Germany.
Thus the Market Cap represents a complementary country wealth measure 
in addition to the GDP. 
The Lorenz curves for Market Cap data \cite{marketcap}
are shown in Fig.~\ref{fig11} in the same presentation style 
as in Fig.~\ref{fig9}
for GDP. As for GDP data we find that the WTH describes well
the real data: certain deviations are present for the RJS model
while the RJE model describes the data almost perfectly.
The parameters of RJS and RJE models are close to those
of the GDP cases in Fig.~\ref{fig9}.
The poverty phase that owns 2\% of total Market Cap 
corresponds to 68\% of countries, 10\% (1\%) of countries own 82\% 
(48\%) of the total Market Cap.
For the RJE model the variation of the poverty and oligarchic
phase with $\varepsilon(RJE)$ (for $a=3.721$, $\varepsilon_C=0.244$) 
are shown in Fig.~\ref{fig12} which is similar to the GDP case in 
Fig.~\ref{fig10}. 

The WTH approach describes well the GDP distributions, wealth 
of country households \cite{piketty2}, and distributions of 
Market Cap of SE companies  presented in \cite{fpuarxiv2025}. 
Thus for three types of players
the WTH theory explains very well the
existing inequality: for wealth of individual persons,
market capitalization of companies and GDP distribution over countries.
The number of persons for each of such a player
is very different. But we argue that
the thermalization takes place for an individual player 
and not for individual persons that belong to a given player.
This situation is similar to a case of a gas in a 3D box composed of 
different atoms: even if the masses are different
for different sorts of atoms still the
average  kinetic energies of atoms are the same
being equal to $3k_B T/2$ where $k_B$ is the Boltzmann constant
and $T$ is the temperature. However, the atom square velocity is
inversely proportional to atomic mass.
Similarly, for the GDP per capita \cite{capita}
the three top positions correspond to 
Monaco, Liechtenstein, Luxembourg
that do not produce a significant influence on the world economy.
Of course, for GDP of countries
certain effects of high or small country population
may play some role but we argue that this does not
affect the RJ thermalization of the GDP distribution.

\begin{figure}[htbp]
\begin{center}
  \includegraphics[width=0.9\textwidth]{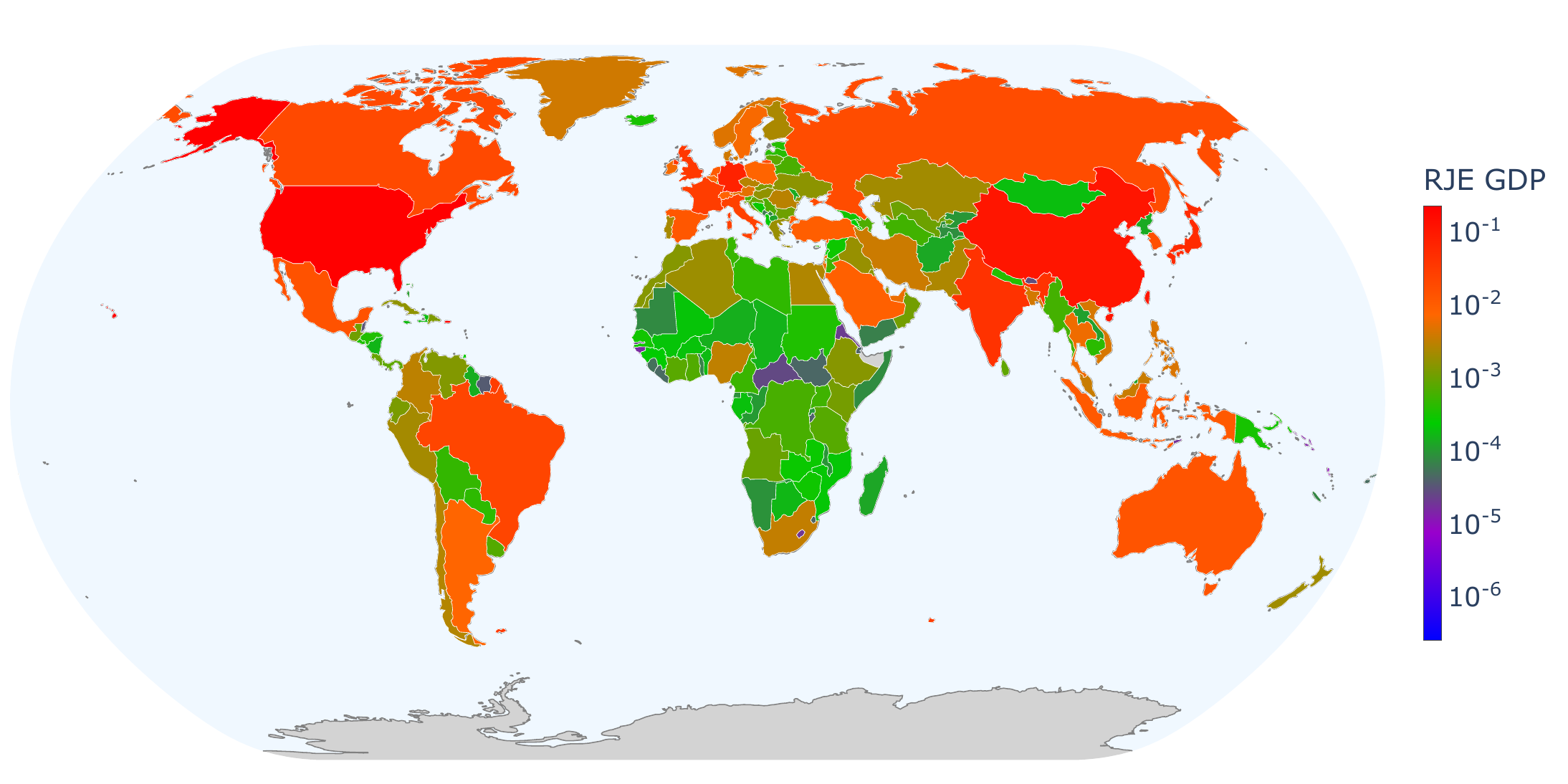}
\caption{\label{fig13}
  World map of RJE GDP values for 212 countries 
  computed from the RJE Lorenz curve of Fig.~\ref{fig9}(b) 
  at $a=4.31$, $\varepsilon =0.01$ 
  corresponding to the best RJE fit of real GDP UN 2023 data \cite{unyears}
  (see text for details). 
  The color values correspond to sum normalized GDP values 
  in logarithmic scale for the color bar. 
}
\end{center}
\end{figure}

In Fig.~\ref{fig13}, we show the world map of RJE GDP values of countries 
computed from the RJE Lorenz curve of Fig.~\ref{fig9}(b) 
with optimal parameters $a=4.31$ and $\eps=0.01$ for the best fit 
with real GDP UN 2023 data from \cite{gdpwiki}. 
Qualitatively, we attribute to each of the 212 countries of 2023 
with index $m=1,\ldots, 212$ ($m=1$ for the country with minimal and 
$m=212$ for the country with maximal GDP) the cumulated household 
interval $[(m-1)dh,m\,dh]$, $dh=1/212$ and take the wealth increase 
$w_m^{\rm (RJE)}=w(m\,dh)-w((m-1)dh)$ in this interval 
using the RJE curve $w(h)$ which provides sum normalized wealth values 
$\sum_m w_m^{\rm (RJE)}=w(1)-w(0)=1$. 
Mathematically, we actually used the formula 
$w_m^{\rm (RJE)}=CE_h((m-0.5)/212)$ where 
the constant $C$ is determined by the sum normalization and 
the function $E_h(h)$ is given by the analytic expression 
$(\ref{eqEh1})$ of Appendix A.2 for the RJE energy expressed as 
a function of cumulated household $h$ 
computed for the limit $N^{\rm (RJE)}\to\infty$. This 
is a non-trivial function since one has to take 
the RJE energy $E_k=(e^{ak/N^{\rm (RJE)}}-1)/(e^a-1)$ 
at the value of the rescaled index $k/N^{\rm (RJE)}$ that corresponds to a 
given household value $h\in[0,1]$ (see Appendix for details). 

A similar world map figure as Fig.~\ref{fig13} but for the real 
(sum normalized) GDP UN 2023 data is shown Appendix Fig.~\ref{figA3}. 
The deviations between the two maps expressed by $\Delta_{GDP}$
are presented in the world map of (relative) differences in 
Appendix Fig.~\ref{figA4}. 
Furthermore, in Appendix Fig.~\ref{figA5}, we also show in logarithmic scale 
the direct dependence of $w_m^{\rm (RJE)}$ and the sum normalized real data 
values $w_m$ on the rescaled household index 
$h=(m-0.5)/212$ for each country. Both curves cover 6 orders of magnitude 
for the rescaled GDP values and are very close but 
for a few countries there are significant deviations. Essentially, 
$w_m^{\rm (RJE)}$ represents a smoothed curve of the real values 
$w_m$. 
In Appendix A.3, we also explain the link between the number of 
interacting agents in the RJE model and the cumulated number of 
households/countries. This number corresponds to the number 
of RJE energy modes $k$ such that the associated cumulated 
household value for this 
mode index falls into the same cumulated household interval for a 
given country. 

\section{WTH Lorenz curves for stock exchange markets}

In this section, we compare the RJE Lorenz and Pareto curves 
to real data of Market CAPitalization (MCAP) of
companies at Hong Kong SE, Shanghai SE and London SE
using publicly available data. 
According to the WTH concept, we assume that interactions
between agents representing companies of a given SE
lead to the RJ thermal distribution (\ref{eqrj}) which allows 
to construct Lorenz and Pareto curves as described in Section 2.

\begin{figure}[htbp]
\begin{center}
\includegraphics[width=0.9\textwidth]{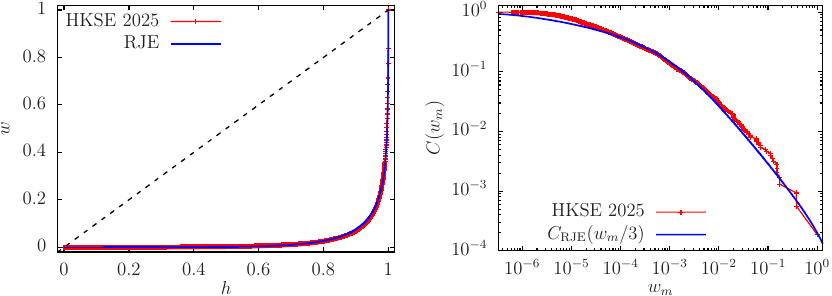}
\caption{\label{fig14}
Lorenz (left panel) and Pareto  (right panel) curves
for MCAP of companies of  Hong Kong stock exchange (HKSE)
at 19 June 2025 \cite{hk} (red curve/symbols)  with $G=0.947$
and the RJE curves (blue curve) with optimal 
$a=7.04$, $\eps=0.000706$ and continuous limit $N \rightarrow \infty$. 
The blue Pareto curve for the RJE model 
is shown with a rescaled argument $C_{\rm RJE}(w_m/3)$ which
excludes points at the very tail of $C(w_m)$
with small statistics (here $w_m$ is rescaled
by the maximal company MCAP); the total number of companies is
$N=2683$.
}
\end{center}
\end{figure}

Thus in Fig.~\ref{fig14}, 
we compare the Hong Kong SE MCAP data 
at 19 June 2025 with the RJE model with parameters obtained by a best fit 
(for the Lorenz curve). The RJE model describes very well not only
the Lorenz curve but also the Pareto curve which shows variations 
of $C(w_m)$ values for almost 4 orders of magnitude.
The data clearly show that the Pareto curve
has a curvature and cannot be described by a simple power law decay
with exponent 1 or 2 that it usually argued to be valid in economy 
studies (see e.g. \cite{yakovenko2}).
In contrast to this algebraic decay of Pareto probability, we show
that the decay is well described by the RJE model
that reproduces the curved Pareto probability decay. The 
approximate RJE expression (\ref{eqCw2main}) agrees with the blue curve 
for $w_m\ge w_{\min}\approx 0.01$.

\begin{figure}[htbp]
\begin{center}
\includegraphics[width=0.9\textwidth]{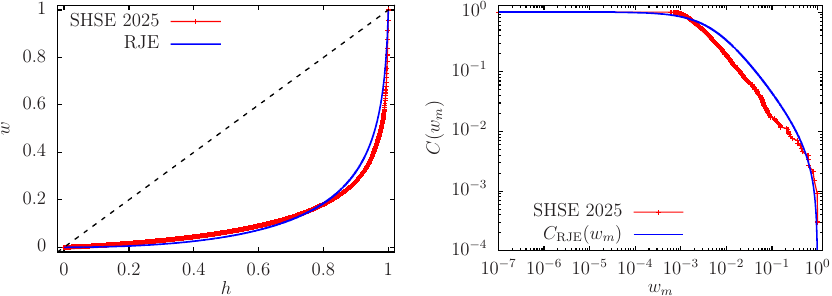}
\caption{\label{fig15}
  Lorenz (left panel) and Pareto  (right panel) curves
  in the same format as Fig.~\ref{fig14}
for the MCAP of companies of the Shanghai SE (SHSE)
at 18 November 2025 with $G=0.778$ and data from 
\cite{sh}
(red symbols, all of them are shown) 
and the RJE model (blue curves) with optimal 
$a=5.27$, $\eps=0.0222$ and continuous limit $N \rightarrow \infty$.
Here the number of companies is $N=1652$.
}
\end{center}
\end{figure}

In Fig.~\ref{fig15} we show Lorenz and Pareto curves for the Shanghai SE
at 18 November 2025 \cite{sh} in the same format as in Fig.~\ref{fig14}.
The RJE model gives a satisfactory description of this data
even if deviations from real data are present for the Pareto curve. 
Here, the 
approximate RJE expression (\ref{eqCw2main}) agrees with the blue curve 
for $w_m\ge w_{\min}\approx 0.05$.

\begin{figure}[htbp]
\begin{center}
\includegraphics[width=0.9\textwidth]{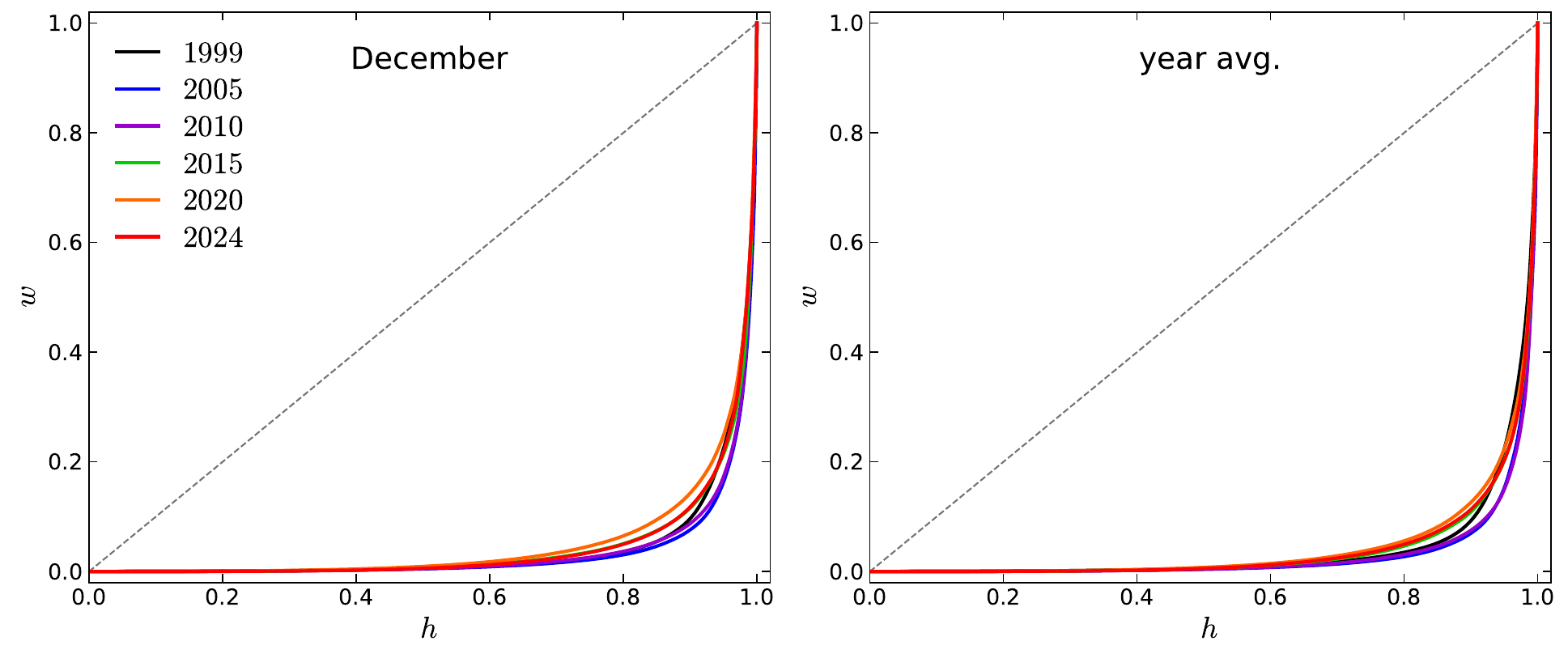}
\caption{\label{fig16}
  Lorenz curves at different years for the London Stock Exchange (LSE) 
  at December (left panel)
  and averaged over the full year (right panel) for 
  the same years as in the left panel..
}
\end{center}
\end{figure}

The London SE (LSE) provides an excellent detailed public archive \cite{london}
of MCAP data for the years 1999 to 2024. There are data for each month
and averaged data for each year. In Fig.~\ref{fig16} we show
the Lorenz curves for 6 selected years with data for December
of a given year and data averaged over the full year.
There is a certain variation of curves but in global
this variation is moderate. Also the Gini coefficient
has rather small variations during these years
being in the range $0.919 \leq G \leq 0.938$
(see data in Table~\ref{tab:inequality}). This Table represents
Gini coefficients and other data for real Lorenz data studied in this work,
we discuss this Table in detail a bit later.

\begin{figure}[htbp]
\begin{center}
\includegraphics[width=0.8\textwidth]{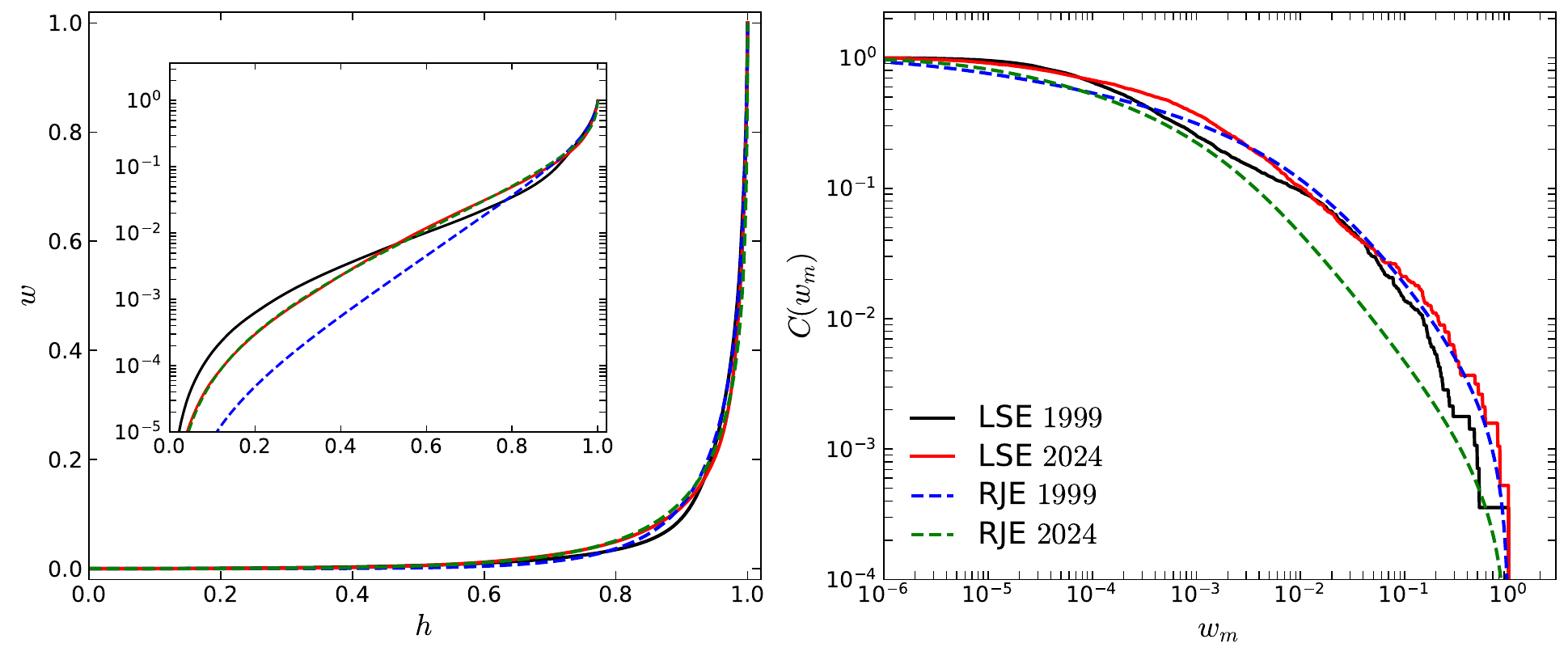}
\caption{\label{fig17}
  Year averaged Lorenz curves for years 1999 and 2024 of the LSE (left panel)
  and Pareto curves for the same data (right panel);
  LSE data are shown by full curves and RJE model data
  are shown by dashed curves; here
  $G=0.920$ in 1999 with $N=2807$ companies
  and $G=0.917$ in 2024 with $N=1900$; RJE parameters are
  $\varepsilon= 0.0087$, $a=3.79$ in 1999; $\varepsilon= 0.0030$, 
  $a=5.64$ in 2024;
  RJE curves are obtained in the limit $N \rightarrow \infty$.
  Insert in left panel shows 
  data in logarithmic scale for the $y$-axis.
}
\end{center}
\end{figure}

The comparison of LSE data with the RJE model for the years 1999 and 2024 
are shown in Fig.~\ref{fig17}. For 2024 the Lorenz curve for real data
(red)  and RJE curve  (dashed green)
are  very close even in logarithmic scale for wealth $w$; 
however, for Pareto curves
the deviations are well visible. For 1999 we have strong deviations
for wealth values $w < 0.01$ while the  Pareto curves, in black and dashed blue,
are closer to each other. In global, we consider that the WTH theory
gives a satisfactory description of the LSE data for these years.
We remind that the RJE parameters $\varepsilon$ and $a$ are
obtained for the Lorenz curve data (in normal scale) and stress more 
moderate values of wealth while the Pareto description stresses much more 
the region of very high wealth values. 
We also consider that the WTH description is not likely to 
be very exact for very small wealth values with $w < 0.01$
since the thermalization process can be very slow in the condensate phase. 
Furthermore, certain deviations between raw data and the RJE curves 
in logarithmic scale for very small values 
of $w$ (Lorenz curves) or $C(w_m)$ (Pareto curves) are to be expected 
since the fit procedure to compute the optimal RJE parameters $a$ and 
$\eps$, by minimizing a certain functional for the geometric curve distance, 
are done for the Lorenz curves in normal scale. Without going into 
much detail, we mention that modified fit procedures in logarithmic scale 
(or for Pareto curves) may provide different values of $a$ and $\eps$ 
such that the corresponding curves agree better in logarithmic scale but 
then the quality of the fit in normal scale is reduced.

\begin{figure}[htbp]
\begin{center}
\includegraphics[width=0.8\textwidth]{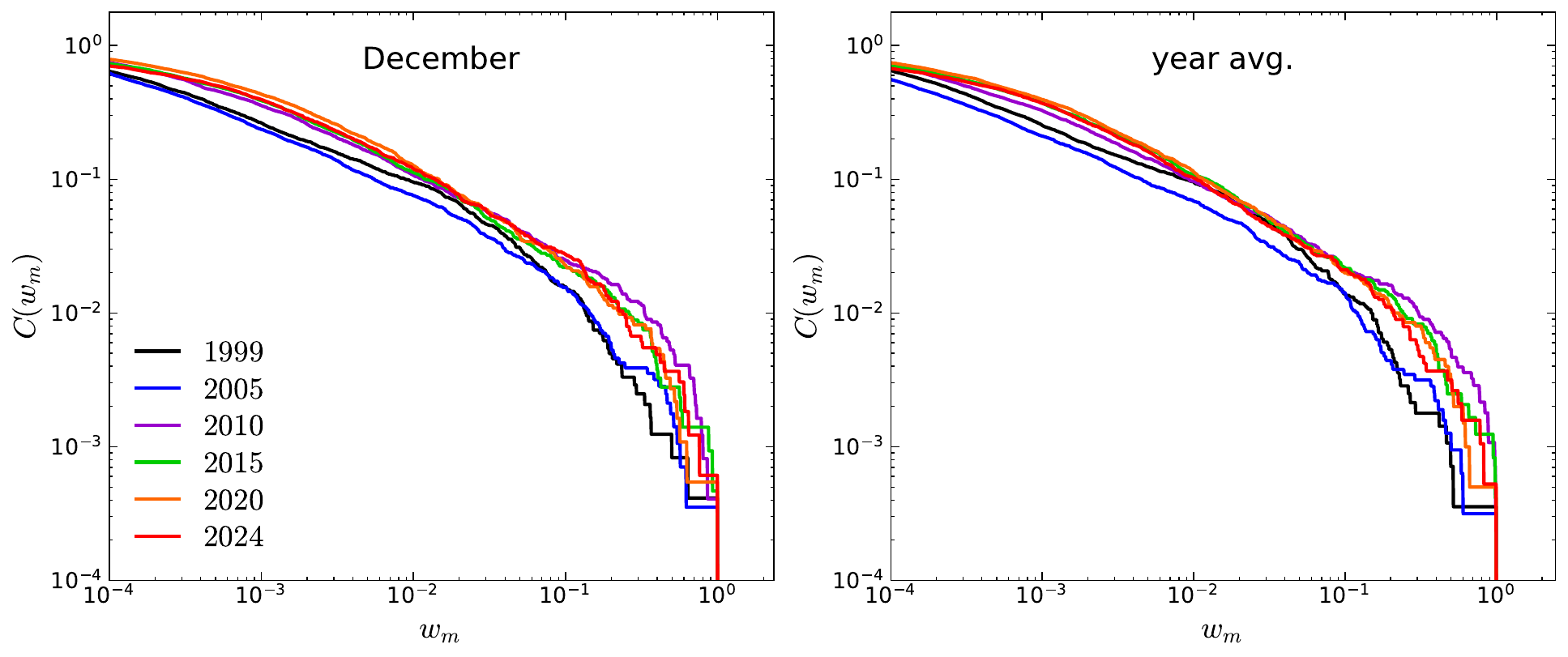}
\caption{\label{fig18}
  Pareto curves for the same LSE data as in Fig.~\ref{fig16}.
}
\end{center}
\end{figure}

In Fig.~\ref{fig18} we present the Pareto curves for the years and data
shown in Fig.~\ref{fig16} with Lorenz curves. This data provides size 
variations by approximately 3-4 orders of magnitude for both 
quantities $C(w_m)$ and $w_m$. We consider
that these Pareto curves cannot be described by a simple algebraic
decay in such a large interval.

\begin{table}[H]
\caption{Wealth inequality indicators for World, US, DE, FR, GDP of countries 
(UN data from 1973, 1993, 2023), Stock Exchange 
(Hong Kong HKSE, Shanghai SHSE, 
London LSE; year averages), the Bitcoin transaction network (quarterly data) 
and World Trade Network (WTN). The columns correspond to system name, 
Gini coefficient $G$, share of total wealth held by the bottom 50\%,
the top 10\%, and the top 1\% of households/countries/companies/wallets. 
\label{tab:inequality}}
\begin{tabular}{lcccc}
\toprule
\textbf{System} & \textbf{Gini} $G$ & \textbf{Bottom 50\%} & \textbf{Top 10\%} & \textbf{Top 1\%} \\
\midrule
\multicolumn{5}{l}{\textit{World and countries --- annual-average household wealth}} \\[2pt]
World 2021 & 0.842 & 2.0\% & 75.3\% & 37.3\% \\
US 2019 & 0.852 & 1.5\% & 76.7\% & 37.4\% \\
DE 2010 & 0.749 & 3.2\% & 59.0\% & 18.3\% \\
FR 2010 & 0.676 & 5.4\% & 49.9\% & 11.7\% \\
\midrule
\multicolumn{5}{l}{\textit{Gross domestic product --- annual-average country GDP (UN data)}} \\[2pt]
GDP UN 1973 & 0.892 & 0.7\% & 84.3\% & 37.2\% \\
GDP UN 1993 & 0.904 & 0.8\% & 86.3\% & 44.3\% \\
GDP UN 2023 & 0.880 & 1.0\% & 82.2\% & 43.8\% \\
\midrule
\multicolumn{5}{l}{\textit{Stock Exchange --- annual-average market capitalization}} \\[2pt]
HKSE 2025 & 0.947 & 0.3\% & 93.2\% & 63.6\% \\
SHSE 2025 & 0.778 & 6.4\% & 72.2\% & 38.8\% \\
LSE 1999 & 0.920 & 0.6\% & 90.7\% & 40.9\% \\
LSE 2005 & 0.938 & 0.4\% & 93.0\% & 50.7\% \\
LSE 2010 & 0.937 & 0.4\% & 92.5\% & 51.0\% \\
LSE 2015 & 0.919 & 0.6\% & 89.1\% & 46.2\% \\
LSE 2020 & 0.909 & 0.7\% & 87.4\% & 44.7\% \\
LSE 2024 & 0.917 & 0.5\% & 88.7\% & 45.8\% \\
\midrule
\multicolumn{5}{l}{\textit{Bitcoin transaction network --- wallet wealth (Q1 snapshots)}} \\[2pt]
Bitcoin 2011 Q1 & 0.887 & 1.6\% & 83.5\% & 54.5\% \\
Bitcoin 2012 Q1 & 0.917 & 0.9\% & 87.5\% & 68.5\% \\
Bitcoin 2013 Q1 & 0.944 & 0.5\% & 92.8\% & 65.8\% \\
\midrule
\multicolumn{5}{l}{\textit{World Trade Network --- country total trade (imports $+$ exports)}} \\[2pt]
WTN 1965 & 0.770 & 3.6\% & 67.3\% & 12.9\% \\
WTN 1975 & 0.802 & 2.3\% & 70.2\% & 16.2\% \\
WTN 1985 & 0.794 & 2.1\% & 68.8\% & 17.2\% \\
WTN 1995 & 0.832 & 1.2\% & 73.1\% & 17.3\% \\
WTN 2005 & 0.845 & 1.0\% & 75.0\% & 19.8\% \\
WTN 2014 & 0.801 & 1.7\% & 65.6\% & 18.0\% \\
\bottomrule
\end{tabular}
\end{table}

Globally, we consider that the WTH theory
provides a good description of MCAP Lorenz and Pareto curves
for Hong Kong, Shanghai and London SE. 

We did not find public data for the New York SE,
while S\&P data is limited to only 500 companies 
representing only a quarter of all NYSE companies. 
For this reason, we do not discuss the NYSE here.

\section{WTH Lorenz curves for bitcoin transactions}

In this Section we present WTH results for  Lorenz and Pareto curves of
bitcoin transactions. The first Lorenz curves for bitcoins
were shown in \cite{bitcoin} (see Fig.~15 there)
on the bases of the transactions collected by Ivan Brugere 
for the period 2009 to April 2013 and publicly available at
\cite{bitcollect}. An interested reader
can find more information about bitcoins e.g. at
\cite{blockchain,shamir,biryukov,bitcoinscience}.
The transactions are done between so called wallets \cite{blockchain},
in \cite{bitcoin} it was shown that ingoing and outgoing transactions
between wallets give almost the same Lorenz curves (see Fig.15 at \cite{bitcoin}),
so that here we present the Lorenz curves using
the sum of ingoing and outgoing transactions.
More details about the properties of the bitcoin network for the above
period are available at \cite{bitcoin}.

\begin{figure}[htbp]
\begin{center}
\includegraphics[width=0.8\textwidth]{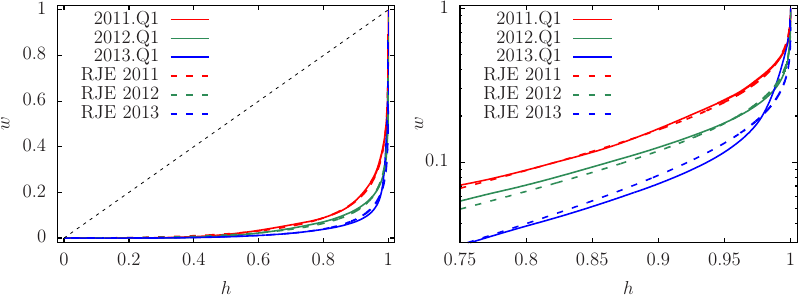}
\caption{\label{fig19}
  Comparison of the Lorenz curves for Bitcoin data in 2011Q1, 2012Q1, 2013Q1
  (full curves) with the corresponding RJE curves with optimal 
  fit parameters for each case (dashed curves). 
  The respective Gini coefficients are 
  $G=0.887, 0.917, 0.944$ and the RJE model parameters are
  $\varepsilon = 1.30\times 10^{-3}, 1.05\times 10^{-4}, 2.64\times 10^{-5}$
  and $a = 7.61, 10.48, 11.43$; left panel shows all scales
  and right panel shows a zoom for rich wallets
  in semi-logarithmic scale. 
}
\end{center}
\end{figure}

Here, we present Lorenz and Pareto curves for bitcoin
transaction collected for the first quarter (Q1) of  2011Q1, 2012Q1, 2013Q1. 
The Lorenz curves shown in Fig.~\ref{fig19} match very well the 
associated RJE curves (with optimal fit parameters given in the caption). 
We note that these cases correspond to quite large values of 
$G\sim 0.9$-$0.95$ and $a\sim 7$-$12$ with very small values for 
$\eps\sim 10^{-3}$ or smaller. 

\begin{figure}[htbp]
\begin{center}
\includegraphics[width=0.8\textwidth]{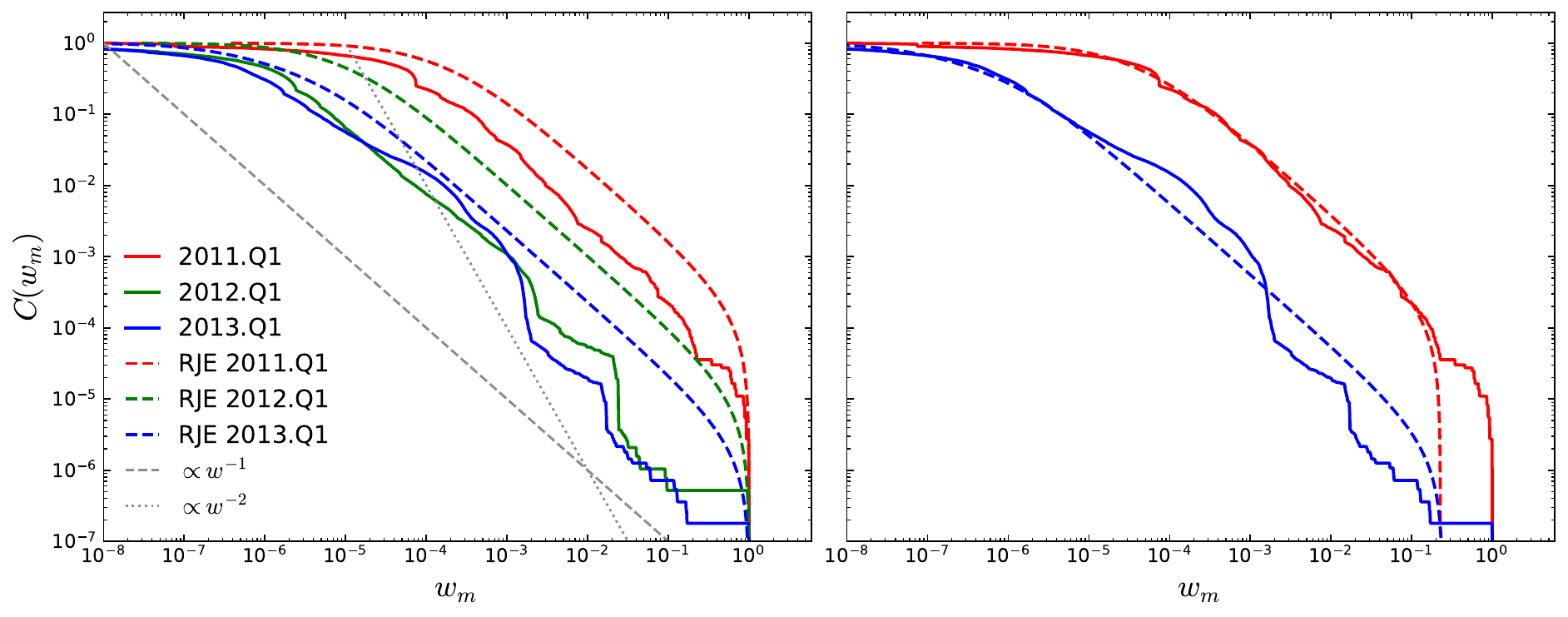}
\caption{\label{fig20}
  Pareto curves for the same Bitcoin data used in Fig.~\ref{fig19} 
  (full curves); RJE curves are shown by dashed lines; 
  left panel shows all cases (with $w_m$ rescaled by its maximal value) and 
  right panel shows data and shifted RJE curves for the cases of 
  2011Q1 and 2013Q1 where the horizontal shift to the left excludes 
  the contribution of $C(w_m)$ data with very low transaction statistics at 
  the very tail.
  The black dashed and dotted straight lines in the left panel show 
  the algebraic slopes $C(w_m)  \propto 1/w_m , 1/w_m^2$.
}
\end{center}
\end{figure}

The comparison of real Pareto curves (with the data sets of Fig.~\ref{fig19})
and the curves of the RJE model 
(with the same parameters as in Fig.~\ref{fig19})
is shown in Fig.~\ref{fig20}. The variation range for the Pareto curve 
is 7 orders of magnitude for $C((w_m)$ and 8 orders of magnitude for $w_m$.
There are significant fluctuations of the Pareto probability
especially visible for the 2012Q1 and 2013Q1 data sets.
However, the RJE model provides a satisfactory description
of the global behavior in this huge range of variation
of Pareto probability. We note that the tail of $C(w_m)$
corresponds to a small transaction statistics (only a few events)
and their exclusion significantly improves the matching with 
the RJE Pareto curves (see right panel of Fig.~\ref{fig20}).

Due to the large values of $a$ and small values of $\eps$ (with even 
smaller values of $|\mu|$) we are in a regime of the RJE model where 
the approximate expression (\ref{eqCw2main}) is valid over a very 
large interval $w_m\in[w_{\min},1]$, e.g. with 
$w_{\min}\approx 3\times 10^{-5}$ 
for the Bitcoin data of 2013.Q1 (see blue full and dashed lines). 
In the reduced interval $w_m\in[w_{\min},0.1]$, we can neglect the constant 
term in (\ref{eqCw2main}) which gives the pure power law behavior 
$C(w_m)\approx C_2\,w_m^{-1}$ with exponent $-1$ and $C_2\approx (\eps-\mu)/a$ 
corresponding to a straight dashed line blue (for this interval). Also 
for the other two cases of  2011.Q1 and 2012.Q1, we can identify a significant 
interval where the same power law is valid (see regions where the red 
and green dashed lines are straight with exponent/slope being $-1$). 
However, for these two cases the interval is slighter smaller 
than for the case of 2013.Q1 with largest $a$ and $G$ values. 
We insist that in all cases the power law does not extend to the 
full interval up to $w_{\max}=1$ and here the constant term in 
(\ref{eqCw2main}) is indeed important since it implies $C(w_m=1)=0$ and a 
curved line in double logarithmic representation (for $w_m\in[0.1,1]$).

Finally, in relation to the statistical description of bitcoin
transactions, we note that there are recent
works \cite{korea1,korea2} where the bitcoin  statistics
is discussed in the framework of the Bose-Einstein thermal distribution
of bosonic particles based on the argument 
that bitcoins are non-distinguishable like bosons. We find this idea somewhat strange
since there are no obvious quantum aspects with bitcoins. Therefore, 
their thermal distribution should be described by the RJ 
distribution (\ref{eqrj})
valid for the classical mechanics and classical objects
which is very different from the quantum BE distribution (\ref{eqbe}).
It is true that at high temperatures the BE distribution is reduced to 
the RJ case 
but at such temperatures there is no RJ condensation
(see above our notes about works
of Fr\"ohlich \cite{frohlich1,frohlich2}).
Also the data sets studied in \cite{korea1,korea2} are significantly smaller
compared to those considered here. 

\section{WTH Lorenz curves for the world trade UN data}

Here we present the WTH description for
the Lorenz curves obtained from the UN COMTRADE data of trade between 
the world countries
studied in \cite{wtn1,wtn2} for the years 1965 to 2014.
During this period the number of countries $N$ varies from
86 in 1965 year up to 158 in year 2014 (maximum of reporting countries
is 169 in 2005). For this period the sum of import and export for all
these countries varies from 0.34 trillions USD up tp 36.83 trillions in 2014.
As for bitcoin transactions the Lorenz curves for import
(ingoing flow) and export (outgoing flow) are very close to each other
and thus we present the effective wealth $w_m$ of a country as a sum
of import and export. 

\begin{figure}[htbp]
\begin{center}
\includegraphics[width=0.8\textwidth]{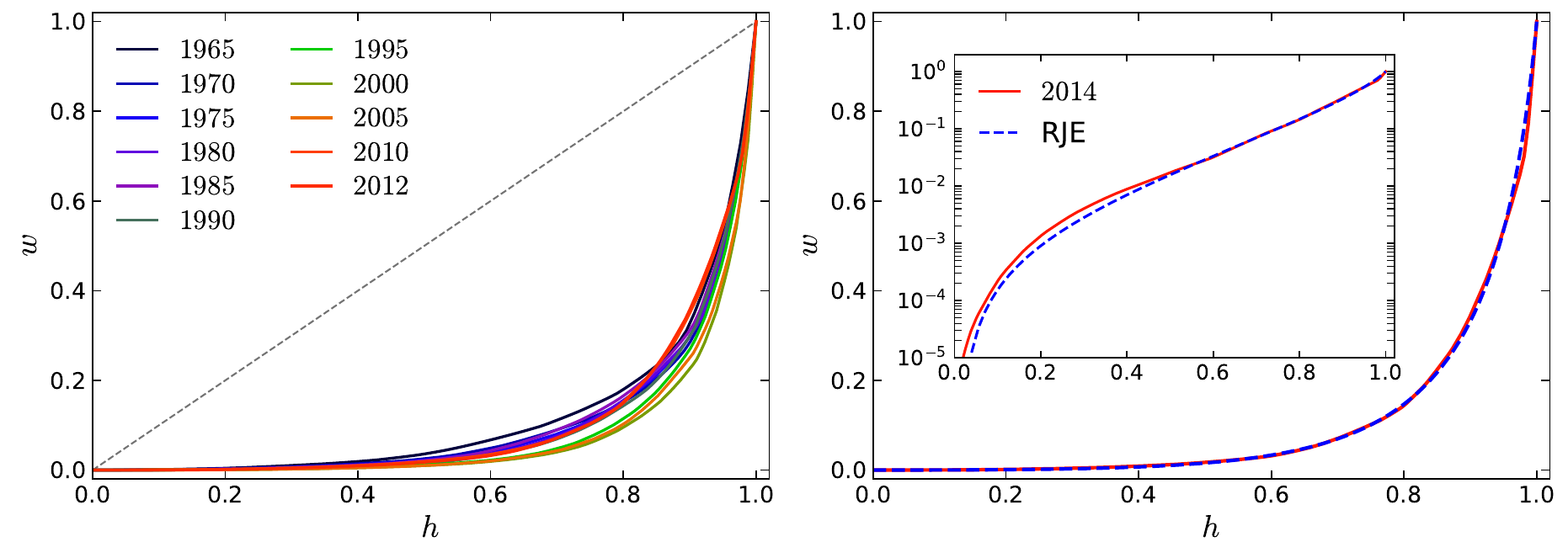}
\caption{\label{fig21}
  {\em Left panel:} Lorenz curves for UN world trade data at different years;
  {\rm Right panel:} comparison of data for 2014 with the RJE Lorenz curve 
  with parameters $\varepsilon = 0.0623$, $a=1.556$; Gini coefficient is
  $G=0.801$ in 2014; other Gini values are given in Table~\ref{tab:inequality}.
  The insert in the right panel shows the Lorenz curves in logarithmic scale 
  for $y$-axis.
}
\end{center}
\end{figure}

The Lorenz curves for the world trade are shown in Fig.~\ref{fig21},
the Gini coefficients and other parameters of
these curves are given in Table~\ref{tab:inequality}.
We see that during 50 years the Lorenz curves and related Gini coefficients 
remain rather stable showing only moderate variations. 
This is similar to the case of GDP Lorenz curves
in Fig.~\ref{fig8}. The comparison of the world trade Lorenz curve 
of 2014 with the RJE model is presented in the right panel of 
Fig.~\ref{fig21}. 
It shows a good agreement between trade data and the RJE model
even in the logarithmic scale for cumulated wealth.

\section{Universality of the thermodynamic description of inequality}

Above we presented analysis of real Lorenz curves for five
types of wealth for five types of societies 
being households of countries and the world,
GDP of world counties, market capitalization of companies
at leading stock exchange of Hong Kong, Shanghai, London, 
bitcoin transactions between wallets and
the world trade between countries.
The characteristics of Lorenz curves are
presented in Table~\ref{tab:inequality}
in a style similar to those used in \cite{piketty2}.
For all types of society, we have the Gini coefficient
being above $G > 0.77$ (except for the households in France and Germany). 
$G$ is especially high for market capitalization
and bitcoin transactions. For all 5 society types we
have a huge poverty phase when 50\% of households
(or players or agents) own 2\% of total wealth or less
(with exceptions of DE, FR in 2010, Shanghai SE, WTN in 1965, 1975).
For almost all SE cases we have even less than 1\% (except for SHSE).
At the same time the oligarchic phase of top 10\%  
owns more than 75\% of total wealth
and those of top 1\%  owns more than 37\% of total wealth
(with some exceptions for DE, FR and WTN). Moreover for the SE cases
and bitcoins the phase of top 10\% owns almost 90\% (with some exceptions).
In our opinion these results 
demonstrate universal features of wealth inequality
in various types of societies and in the whole world.

The thermodynamics is a universal physical concept \cite{landau}
that describes a variety of physical systems including 
the black-body cosmic microwave background in the Universe
described by the quantum BE distribution \cite{fixsen,wikicmb}.
For classical systems the BE distribution (\ref{eqbe}) is replaced 
by the RJ distribution (\ref{eqrj}) 
with temperature and chemical potential
appearing due to energy and probability norm conservation.
Above, we presented the RJ thermodynamic description for
the wealth distributions in all 5 types of societies
showing that the WTH theory gives a good description
of real Lorenz curves and also Pareto curves
(for households of countries,
companies at SE and bitcoin wallets when the
number of agents is large).
Thus we argue that the thermodynamic theory
well describes the wealth inequality in the world.
The most important feature of the RJ description
is that within this concept the condensation of
poor households appears very naturally at
low relative energies/wealth of the system.
We note that similar constraint driven
condensation appears also in a variety of physical systems
(see e.g. \cite{trizac,satya,marsili}.

\section{Discussion and conclusion}

In this work, we described the thermodynamic properties of
classical systems with two integrals of motion,
being energy and probability norm,
where the RJ distribution has a condensation phase
of low energy states. We argue that energy can be associated with
wealth of agents in social systems
where, according to the  WTH proposed in \cite{fpuarxiv2025},
the RJ thermalization naturally leads to
emergence of a huge poverty phase that owns only a few percents
of total wealth and a tiny oligarchic phase
that contains a huge amount of the total wealth.
We presented real Lorenz curves of five types of societies
(households of countries,
GDP of countries, companies of stock exchange,
bitcoin wallets and world trade between countries)
and demonstrated that the RJ description depicts well these Lorenz curves.
The WTH RJ theory also gives
a good description of Pareto curves
for the cases with many agents (households, companies, wallets).
The RJ thermalization can appear due to
nonlinear interactions between agents leading to dynamical chaos
as it is shown in \cite{fpuarxiv2025,sssarxiv}
for social networks with social stratification.

We note that it is  possible to decrease the poverty phase
by a reduction of wealth dispersion $B$ in a society
that leads to an increase of relative system energy $\varepsilon$
as it is shown in Fig.~\ref{fig10}. 
Such an action reminds the Zucman tax for high revenues
(see e.g. \cite{zucman1,zucman2}).
From the mathematical description of the RJ thermalization
such an action would indeed reduce the dispersion $B$
and increase relative system energy $\varepsilon$
thus reducing the wealth inequality. 
However, we leave to economists and sociologists
a discussion of effects of such an action in
real human society.

\section*{Appendix}
\setcounter{equation}{0}
\renewcommand{\theequation}{A\arabic{equation}}
\setcounter{figure}{0}
\setcounter{section}{0}
\renewcommand\thefigure{A\arabic{figure}}
\renewcommand\thesection{A\arabic{section}}
\renewcommand{\figurename}{Appendix Figure}

\section{Analytical results for the continuous RJS model}

\subsection{Continuous limit and chemical potential}

For the RJS and RJE model with finite $\varepsilon$ it is 
possible to obtain explicit formulas in the continuous 
limit $N\to\infty$ by replacing the sums over $m$ with integrals 
over an rescaled index variable $t=m/N\in[0,1]$. In this section, 
we will present some results on this for the RJS model and the 
extension to the RJE model will be given in the next appendix section. 

For the RJS model, we have $E_m=m/N=t$ and in the following, we also 
use $\eps=E$ (since $E_0=0$ and $B=(N-1)/N\to 1$ for $N\to\infty$).
In the limit $N\to\infty$ the implicit equation (\ref{eqS3}) becomes:
\begin{align}
\label{eqmuA} 
1&=(\eps-\mu)\int_0^1 \frac{1}{t-\mu}\,dt
=(\eps-\mu)\ln\left(\frac{1-\mu}{-\mu}\right)
\end{align}
which can be rewritten in the following form:
\begin{align}
\label{eqmu2} 
\mu=-(1-\mu)e^{-1/(\eps-\mu)} \; .
\end{align}
Both equations determine $\mu$ as a function of $\eps\in]0,1[$. 
In the limit of small $\eps$ one can simply iterate Eq.~(\ref{eqmu2}) by 
inserting $\mu_0=0$ in the RHS which gives $\mu_1=-e^{-1/\eps}$ on the LHS 
which can be inserted in the RHS to obtain a better value $\mu_2$ etc. 
This procedure converges nicely for small $\eps$ and for other values 
of $\eps$ one 
can use standard techniques to solve these 
equations numerically and efficiently. For $\eps \ll 1$, the 
first approximation $(-\mu)\approx e^{-1/\eps}\ll \eps$ is already very good.

To understand the limit of $|\mu|\gg 1$ it is more useful to consider 
$\eps$ as a function of $\mu$ which is determined by (\ref{eqmuA}). 
Expanding the logarithm in (\ref{eqmuA}) up to 3rd order in $1/\mu$ one 
finds that 
\begin{align}
\label{eqmuinfty}
\eps\approx \frac12\left(1+\frac{1}{6\mu}\right)\to \frac12
\end{align}
for $|\mu|\to\infty$ 
which is expected from the curve of $\mu$ in Fig.~\ref{fig1}. The $1/\mu$ 
correction in (\ref{eqmuinfty}) will be useful below. 

\subsection{Lorenz curve}

As explained in the main part of this work, to compute the Lorenz curve 
we have to compute a partial sum over 
$\rho_m=(E-\mu)/[N(E_m-\mu]\to (\eps-\mu)/(t-\mu)\,dt$ (with 
$dt=1/N$ and $t=E_m=m/N$) to obtain the household 
fraction $h$ and over $(E_m/\eps)\rho_m$ to obtain the wealth variable. 
Now, we replace these partial sums also by integrals up to some 
arbitrary value $s\in[0,1]$ which provides functions $h(s)$ and $w(s)$ 
allowing to determine the Lorenz curve $w(h)$. These partial integrals are:
\begin{align}
h(s)&=(\eps-\mu)\int_0^s \frac{1}{t-\mu}\,dt
=(\eps-\mu)\ln\left(\frac{s-\mu}{-\mu}\right) 
\label{sofh}
\folgt s(h)=(-\mu)\left(e^{h/(\eps-\mu)}-1\right)
\end{align}
and
\begin{align}
w(s)&=\frac{\eps-\mu}{\eps}\int_0^s \frac{t}{t-\mu}\,dt
=\frac{\eps-\mu}{\eps}\int_0^s \left(1+\frac{\mu}{t-\mu}\right)\,dt
=\frac{\eps-\mu}{\eps}s+\frac{\mu}{\eps}h(s)
\label{eqwofs}
\ .
\end{align}
Inserting (the 2nd expression of) (\ref{sofh}) in (\ref{eqwofs}) we obtain 
the following analytical expressions for the Lorenz curve:
\begin{align}
w(h)&=\frac{-\mu}{\eps}\left(
(\eps-\mu)\left(e^{h/(\eps-\mu)}-1\right)-h\right) 
\label{eqlorA}
=\frac{1-\mu}{\eps}\,e^{-1/(\eps-\mu)}\left(
(\eps-\mu)\left(e^{h/(\eps-\mu)}-1\right)-h\right)\ .
\end{align}
Here the 2nd expression of (\ref{eqlorA}) has been obtained by replacing 
the global factor 
$\mu$ with (\ref{eqmu2}) which gives a more convenient expression. 
Using (\ref{eqmu2}), one can verify that (\ref{eqlorA}) satisfies 
the conditions $w(0)=0$ and $w(1)=1$. 

The (2nd) expression (\ref{eqlorA}) allows 
to take the limit $\eps\ll 1$ with $\mu\approx -e^{-1/\eps}\ll \eps$ such 
that for $\eps\ll 1$ we have the simplified Lorenz curve (replacing $\mu=0$ 
in (\ref{eqlorA})): 
\begin{align}
\label{eqlorA0}
w(h)&\approx
e^{-1/\eps}\left(e^{h/\eps}-1-\frac{h}{\eps}\right)\approx e^{(h-1)/\eps}\ .
\end{align}
Here both expressions are equivalent approximations for small $\eps$ with 
$e^{-1/\eps}\ll 1$. The first (second) expression does not exactly verify the 
condition for $w(1)$ (or $w(0)$). The second expression is very simple 
and numerically quite sufficient for $\eps\le 0.2$. 

We have verified that both expressions (\ref{eqlorA}) 
coincide with the numerical data shown in the left panel of 
Fig.~\ref{fig4}  up to graphical precision 
with an error below $2\times 10^{-4}$ and for all values of $\eps$ used in Fig.~\ref{fig4}. 
The approximate formulas (\ref{eqlorA0}) 
are valid for $\eps\le 0.2$ with an error $\sim 10^{-2}$ for $\eps=0.2$ (and 
smaller errors for smaller values of $\eps$). 
This can be seen in Fig.~\ref{figA1} comparing the data for $\eps=0.1,0.2,0.3$ 
between the analytic expressions and the data for $N=10000$. Even 
for $\eps=0.3$ only a modest deviation of the approximate curve is visible 
while here and in all other cases the more precise expression (\ref{eqlorA}) 
matches the numerical data very closely. 

\subsection{Other quantities}

Using the analytical expressions for $w(h)$ one can compute several other 
quantities. For example it is interesting to consider the 2nd order expansion 
in $h$ for $|h/(\eps-\mu)|\ll 1$ which gives:
\begin{align}
\label{eqlor2}
w(h)&=\frac{(-\mu)}{2\eps(\eps-\mu)}\,h^2\ .
\end{align}
We know that the limit $|\mu|\to\infty $ corresponds to $\eps\to 1/2$ 
and in this case (\ref{eqlor2}) is valid for all $h\in[0,1]$. This gives 
the very simple formula $w=h^2$ (which is also 
obvious from the fact that $\rho_m=1/N=$ const. for $|\mu|\to\infty$ and 
the way the Lorenz curve is constructed from $\rho_m$). 

\begin{figure}[htbp]
\begin{center}
\includegraphics[width=0.65\textwidth]{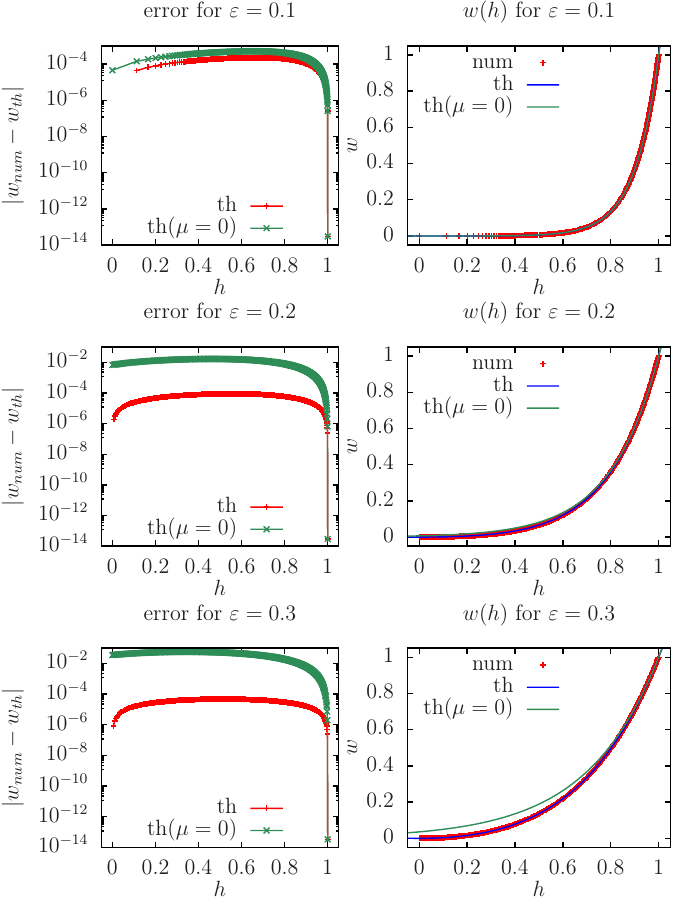}
\caption{\label{figA1}
Comparison of Lorenz curves of cumulated wealth fraction $w$ 
versus cumulated household fraction $h$ for the analytical model with 
the numerical data of the RJS model for finite $N=10000$ and 
for three key values of the rescaled energy 
$\eps=0.1,0.2,0.3$ (top to bottom). 
Left panels shows the difference between the analytical model 
and numerical data and right panels show 
directly the curves $w$ versus $h$ 
for the numerical data (red lines and data points) and the analytical model. 
Blue lines (red data points in left panel) 
correspond to the formula (\ref{eqlorA}) valid 
for all values of $\eps$ and using the appropriate value of the 
chemical potential $\mu$ determined by the implicit equation (\ref{eqmuA}). 
Green lines (green data points in left panel) 
correspond to the (second) approximate formula  
(\ref{eqlorA0}) valid for small $\eps\le 0.2$. 
The discrete points of data in the top right panel for $\eps=0.1$ at 
values close to $w=0$ indicate finite values for $\rho_0=0.1129$, 
$\rho_0+\rho_1=0.1660$, etc. which are due to RJ condensation. 
}
\end{center}
\end{figure}

It is also possible to compute the Gini coefficient:
\begin{align}
G&=1-2\int_0^1 w(h)\,dh
=1+\frac{2\mu}{\eps}\left[
(\eps-\mu)^2(e^{1/(\eps-\mu)}-1)-(\eps-\mu)-\frac12
\right]
\label{eqGini2}
=1-\frac{\mu}{\eps}-2(\eps-\mu)\ .
\end{align}
Here the last simpler expression (\ref{eqGini2}) has been obtained 
by replacing the exponential term 
using the implicit 
equation of $\mu$. The limit $\eps\ll 1$ with $\mu\approx -e^{-1/\eps}$ 
gives $G\approx 1-2\eps$ which matches well the values of $G$ given in the 
caption of Fig.~\ref{fig4} for $a=0$ and $\eps\le 0.1$ (rather close 
value for $\eps=0.2$). The 
other values are matched exactly by the more precise expression 
(\ref{eqGini2}). 
Furthermore, inserting the expression (\ref{eqmuinfty}) for large $|\mu|$ 
in (\ref{eqGini2}) one finds (confirms) that $G=1/3$ for $\eps=1/2$ 
(here it is necessary to keep the $1/\mu$ correction in (\ref{eqmuinfty}) 
to obtain the correct result for $G$).

\subsection{General discussion of the continuous limit}

One might be concerned that the integral approximation is not very 
good for small $\mu$ (close to the singularity of the first term in 
(\ref{eqS3})) and some finite but large value of $N$ such as $N=10000$. This 
is true but the integral provides a modified logarithmic singularity which 
allows also to mimic correctly the condensation effect with correct 
probabilities. Therefore even though the values of $\mu$ are modified 
for $\eps\ll 1$ (but still $0<-\mu\ll \eps\ll 1$ for both models !) 
the resulting probabilities (e.g. integrals or sums of $\rho_m$ over 
some interval in $t=m/N$) are the same. The values of $\mu$ obtained 
by the continuous analytical model match very well the curve shown 
in Fig.~\ref{fig1} but of course this figure does not allow to verify if 
$\mu\approx -e^{-1/\eps}$ (continuous model) or $\mu\approx -\eps/(N-1)$ 
(for the finite $N$ model with discrete sums) which are both close to zero 
in the figure. 
In any case, we find that the analytical expressions given here 
(if $\mu$ is properly evaluated by its implicit equation (\ref{eqmuA}) 
and if properly evaluated by avoiding numerical instabilities of some 
formulas in some special cases) match the numerical data with an 
error that scales with $1/N$.

Without going into details, we mention that 
we have also considered a more refined version 
of the continuous model using a finite value of $N$ and keeping the 
first singular term separate from 
the integral (which starts at $s=1/N$ and not $s=0$). In this case, we obtain 
a modified implicit equation of $\mu$ which results in 
values of $\mu$ closer to the model of finite $N$ but the resulting physical 
quantities ($w(h)$ curves, Gini coefficients etc.) are (numerically 
with an error $<10^{-4}$) the same as both 
the numerical data and the simple model. The resulting 
analytical expressions of the refined model 
are slightly modified (essentially replacing $h$ by 
$h-\rho_0$ for $h\ge \rho_0$ 
in the formula of the Lorenz curve and using $w(h)=0$ for $h<\rho_0$ 
where $\rho_0$ may now have a finite value). Note that the 
initial interval 
$h\in[0,\rho_0[$ with exactly $w(h)=0$ for the refined and also the 
discrete model translates to exponentially small values 
$w(h)\approx h^2\,e^{-1/\eps}/(2\eps)$ 
for the simple analytical model (replacing $\mu\approx -e^{-1/\eps}$ in 
(\ref{eqlor2})). 

\section{Analytical results for the continuous RJE model}

\subsection{Lorenz curve}

Here, we present an extension of the continuous limit $N\to\infty$ to the 
RJE model. For this, we consider a real parameter $a$ and 
the (rescaled) RJE spectrum $E_m=E_s(m/N)$ with the 
smooth function $E_s(s)=(e^{as}-1)/D(a)$ where $s\in[0,1]$ is a 
rescaled index variable and $D(a)=e^a-1$ is a parameter dependent on 
$a$ which we will use in this appendix. In this way, the band width 
is again unity $B=E_{N-1}\approx 1$ (and $B=1$ exactly for $N\to\infty$). 
The limit $a\to 0$ is well defined and in this case, we simply recover 
$E_s(s)=s\folgt E_m=m/N$ which is the RJS model as a special case. Negative 
values of $a$ are also possible but less relevant for comparison with real 
data. For the cases where the optimal RJE-fit for some real data 
gives (modest) negative values for $a$, the simple RJS curve is 
typically already a very good fit. 

Now, extending (\ref{sofh}), we have the cumulated household $h(s)$ as a 
a function of the index variable $s$ given by:
\begin{align}
\label{eqRJEhs1}
h(s)&=(\eps-\mu)\int_0^s \frac{1}{E_s(t)-\mu}\,dt
=(\eps-\mu)\int_0^s \frac{D(a)}{e^{at}-b}\,dt
=(\eps-\mu)D(a)\int_0^s \frac{e^{-at}}{1-be^{-at}}\,dt
\end{align}
where we have defined the quantity $b=1+\mu D(a)$ for shorter notations 
and $\mu$ is the chemical potential (typically $\mu<0$ for $T>0$; its 
computation will be discussed in the next subsection). 
The last integral in (\ref{eqRJEhs1}) can be computed in closed form and 
gives:
\begin{align}
\label{eqRJEhs2}
h(s)&
=\frac{(\eps-\mu)D(a)}{ab}\ln\left(\frac{1-be^{-as}}{1-b}\right)\ .
\end{align}
The integral for the cumulated wealth function $w(s)$ (depending 
on the index variable $s$) can be expressed in a similar way as 
(\ref{eqwofs}) for the RJS model:
\begin{align}
w(s)&=\frac{\eps-\mu}{\eps}\int_0^s \frac{E_s(t)}{E_s(t)-\mu}\,dt
=\frac{\eps-\mu}{\eps}\int_0^s \left(1+\frac{\mu}{E_s(t)-\mu}\right)\,dt
=\frac{\eps-\mu}{\eps}s+\frac{\mu}{\eps}h(s)
\label{eqRJEwofs}
\ .
\end{align}
To obtain the Lorenz curve $w(h)$ (cumulated wealth as a function 
of cumulated household), we need to use (\ref{eqRJEhs2}) to express 
$s=s(h)$ as a function of $h$ and insert the result for $s$ in 
(\ref{eqRJEwofs}). This calculation gives:
\begin{align}
\label{eqsh1}
e^{-as(h)}&=\frac{1}{b}
\left[1-(1-b)\exp\left(\frac{abh}{D(a)(\eps-\mu)}\right)\right]
=1+\frac{\mu D(a)}b
\left[\exp\left(\frac{abh}{D(a)(\eps-\mu)}\right)-1\right]
\folgt\\
s(h)&=-\frac{1}{a}\ln\left\{1+\frac{\mu D(a)}b
\left[\exp\left(\frac{abh}{D(a)(\eps-\mu)}\right)-1\right]
\right\}
\quad,\quad b=1+\mu D(a)
\ .
\label{eqsh2}
\end{align}
Using this expression for $s(h)$, we have the closed formula for 
the Lorenz curve:
\begin{align}
w(h)&
=\frac{\eps-\mu}{\eps}s(h)+\frac{\mu}{\eps}h
\label{eqRJEwofh}\ .
\end{align}
We note that (\ref{eqsh2}) 
is written in a way that is numerically stable if the appropriate 
functions \texttt{expm1(x)} for $\exp(x)-1$ and \texttt{log1p(x)} 
for $\ln(1+x)$ are used which are accurate also for $|x|\ll 1$ and 
typically available in any numerical computer library 
(C, Fortran, python, etc.). This point is important since we want 
these expressions also to be useful in the limit of small $a$ or $-\mu$. 
Actually, taking the limit $a\to 0$ (with $D(a)\to 0$ but $a/D(a)\to 1$) one 
obtains from (\ref{eqsh2}) and (\ref{eqRJEwofh}) the first expression 
of (\ref{eqlorA}) for the RJS case as it should be. 

\begin{figure}[htbp]
\begin{center}
\includegraphics[width=0.65\textwidth]{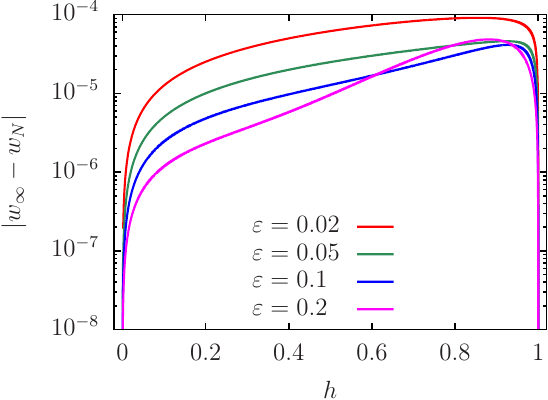}
\caption{\label{figA2}
Difference of Lorenz curves for the RJE model at $a=4.5$ between 
finite $N=10000$ and the continuous limit $N\to\infty$ for the cases 
of the right panel of Fig.~\ref{fig4}. $w_{\infty}$ is 
obtained by the combined expressions (\ref{eqsh2}) and (\ref{eqRJEwofh}) 
while the curve for $w_N$ is obtained for the finite $N=10000$ RJE spectrum 
as described in Section 2.2.
}
\end{center}
\end{figure}
Fig.~\ref{figA2} shows for $a=4.5$ and four $\eps$ values (cases 
of right panel of Fig.~\ref{fig4}) the difference of the Lorenz curves 
obtained 
by the analytic expressions (\ref{eqsh2}), (\ref{eqRJEwofh}) with 
the finite $N=10000$ Lorenz curve constructed as explained in 
Section 2.2. The difference for all cases is clearly below $10^{-4}$. 
(Concerning the computation of $\mu$ see the next subsection.)

\subsection{Implicit computation of the chemical potential}

The above results depend on the value of the chemical potential which is 
determined by the condition $h(s=1)=1$. Using (\ref{eqRJEhs2}) this gives:
\begin{align}
\label{eqmu1}
1&
=\frac{(\eps-\mu)D(a)}{ab}\ln\left(\frac{1-(1+\mu D(a))e^{-a}}{-\mu D(a)}
\right)
=\frac{\eps-\mu}{a(D^{-1}(a)+\mu)}\ln\left(
\frac{ e^{-a}(1-\mu) }{-\mu}\right)\ .
\end{align}
To solve this equation numerically, it is more convenient to reformulate 
it as:
\begin{align}
\label{eqmuB}
\mu&=- e^{-a}(1-\mu) \exp\left(-\frac{a(D^{-1}(a)+\mu)}{\eps-\mu}\right)
=-\frac{1}{\exp\left(\frac{a(D^{-1}(a)+\eps)}{\eps-\mu}\right)-1}\ .
\end{align}
Both variants give a possible fix point iteration, with a first 
approximation by inserting $\mu=0$ in the r.h.s.:
\begin{align}
\label{eqmu000}
\mu\approx -e^{-a}\exp\left(-\frac{a D^{-1}(a)}{\eps}\right)
\quad,\quad\eps\ll 1\ ,
\end{align}
but this fix point iteration 
has a very limited range of convergence which works only for extremely 
small values of $\eps\ll 1$ (depending on $a$). 
However, the approximate value (\ref{eqmu000}) gives a good idea 
of the behavior of $\mu$ at small $\eps$. 
For $a\to 0$, we recover $\mu\approx -e^{-1/\eps}$ already known from 
the RJS model. For large $a$ there are two regimes of 
$\eps\ll a/D(a)$ with $-\mu\sim e^{-(\ldots)/\eps}$ and 
$a/D(a)\ll \eps\ll 1$ with $-\mu \approx e^{-a}\ll 1$. Using 
(\ref{eqmu000}) it is possible to obtain more explicit formulas 
at $\eps\ll 1$ for the Lorenz curve and 
other quantities (computed below) but due to the complications 
for the two different regimes and for lack of space, we do not 
enter into more details on this point. 

Concerning, a careful numerical solution of (\ref{eqmuB}), it turns 
out that there is a technical complication since (\ref{eqmuB}) has 
also a 2nd non-physical solution $\mu=-D^{-1}(a)$ 
(which gives a singular result 
in (\ref{eqsh2}) since it has $b=0$) 
and one must be careful to determine the other physical 
solution with $\mu\neq -D^{-1}(a)$. 
Fortunately, there is a nice numerical trick which consists 
of computing the solution of $0=F_0(\mu)=(\mu-F_1(\mu))/(\mu+D^{-1}(a))$ 
where the denominator $\mu+D^{-1}(a)$ removes the non-physical solution. 
Here $F_1(\mu)$ is either one of the expressions 
on the r.h.s. of (\ref{eqmuB}). In this case, the secant method 
(together with some pre-iterations to find two 
good start values with a sign change) allows to find efficiently 
the zero of $F_0(\mu)$ which is indeed the physical solution. We mention 
that the non-physical solution is not a solution of the initial 
equation (\ref{eqmu1}). However, the latter is numerically quite problematic 
to compute $\mu$. 

We also mention that one can solve approximately the (finite $N$) 
implicit equation (\ref{eqS3}) in the limit of $|\mu|\to\infty$ which shows 
that $\mu<E_0=0$ (with $T>0$) if $E<\sum_m E_m/N$ and 
$\mu>E_{N-1}=B$ (with $T<0$) if $E>\sum_m E_m/N$. For the continuous 
RJE model this means that the transition from $T>0$ to $T<0$ happens 
at the critical energy 
\begin{align}
\eps_C=\frac{1}{N}\sum_{m=0}^{N-1} E_m\to 
\frac{1}{D(a)}\int_0^1 ds (e^{as}-1)=\frac{1}{a}-\frac{1}{e^a-1}
\end{align}
which has the limit $\frac12$ for $a\to 0$ and 
takes the value $\eps_C=0.211$ for $a=4.5$ used in the 
data of Figs.~\ref{figA2} and \ref{fig4} (right panel). 

The analytic results presented so far (and below in the next subsections) 
also apply to the case of negative temperature with $\eps>\eps_C$ and 
$\mu>1$, eventually with some additional complications for the numerical 
stability (especially for $T<0$ with $T\nearrow 0$ and $\mu\searrow B$). 
However, in this work, we do not insist on the case $T<0$ and we typically 
assume $\eps<\eps_C$ with $T>0$ and all optimal RJE fits presented here 
for real raw data correspond to this case.

\subsection{Gini coefficient}

\def\Li{{\rm Li}_2}

Using the analytic result for the Lorenz curve, we can also compute the 
Gini coefficient $G$ by:
\begin{align}
G&=1-2\int_0^1 w(h)dh
=1-\frac{2}{a}\left(1-\frac{\mu}{\eps}\right)F(x,y)-\frac{\mu}{\eps}
\label{eqG1}
\end{align}
where 
\begin{align}
\label{eqFxy}
F(x,y)=-\int_0^1 dh \ln\left(\frac{1-xe^{yh}}{1-x}\right)
\end{align}
with two auxiliary quantities 
\begin{align}
\label{aux}
x=-\mu D(a)>0\quad\mbox{and}
\quad y=\frac{a(1+\mu D(a))}{D(a)(\eps-\mu)}
=\frac{a(1-x)}{D(a)\eps+x}
\ .\end{align}
The integral (\ref{eqFxy}) can be computed in closed form using the 
{\em dilogarithm} defined by:
\begin{align}
\label{eqdilog}
\Li(z)
=-\int_0^z \frac{\ln(1-t)}{t} dt
=\sum_{n=0}^\infty \frac{z^n}{n^2}\ .
\end{align}
Here the first integral expression is well defined for $z\in]-\infty,1]$ 
(maximal real interval where $\Li(z)$ has real values) 
and the power series converges for $|z|\le 1$ (but with 
slow convergence if $|z|\to 1$). The numerical evaluation of $\Li(z)$ can be done very efficiently 
by the power series for $|z|\le \frac12$ with very fast convergence 
and for other arguments $z\in]\frac12,1]$, $z\in[-1,-\frac12[$ 
or $z\in]-\infty,-1[$ 
it is possible to use certain mathematical identities (e.g. reflection 
formula etc., see for example the Wikipedia article on $\Li(z)$) 
to transform the argument inside an interval of nice convergence. 
In the mathematical standard literature, one finds also a different 
very efficient expansion of $\Li(z)=\sum_{n=0}^\infty B_n(-\ln(1-z))^n/(n+1)!$ 
using the Bernoulli numbers $B_n$ with faster 
convergence than the standard series given above (for $|z|\le \frac12$).

Using the dilogarithm (\ref{eqdilog}), we obtain for the function $F(x,y)$ 
the following closed result:
\begin{align}
\label{eqFxy1}
F(x,y)&=\ln(1-x)+\frac{1}{y}\Big(\Li(xe^y)-\Li(x)\Big)\quad\mbox{for}\quad
x<1\quad,\\
\label{eqFxy2}
F(x,y)&=-\frac{y}{2}+\ln\left(1-\frac1x\right)
-\frac{1}{y}\left[\Li\left(\frac1{xe^y}\right)-
\Li\left(\frac1x\right)\right]\quad\mbox{for}\quad
x>1\quad,\\
F(1,0)&=
\lim_{x\to 1} F(x,y(x))=\frac{1}{A}(1-A) \,\ln(1-A)+1
\quad,\quad A=\frac{a}{D(a)\eps+1}
\quad\mbox{for}\quad x=1\quad.
\end{align}
These expressions provide together with (\ref{eqG1}) a closed expression 
for the Gini coefficient $G$. As for the Lorenz curve, we can perform 
the limit $a\to 0$ (which corresponds to the case $x\to 0$) which confirms 
perfectly the (first) expression in (\ref{eqGini2}) for the RJS model 
at $a=0$. Note that the case of $x<0$ is also possible if $a<0$ and the 
above expression (\ref{eqFxy1}) is also valid for this case 
(if $\Li(z)$ is correctly evaluated for $z<0$). 

We have also verified the validity of the analytic result of $G$ by 
a numerical computation of the integral of $w(h)$ over 
the interval $h\in[0,1]$ for 
several values of $a$ and $\eps$. Note that standard numerical integration 
methods such as Simpson or Romberg applied to $w(h)$ (using 
(\ref{eqsh2}) and (\ref{eqRJEwofh})) 
work better, i.e. provide more precise results at given number of 
numerical support points, if the integration is done over the transformed 
variable $u\in[0,1]$ with $h=1-u^4$ (and $dh=-4u^3\,du$) such that 
the difficult region of $h\to 1$ (where $w'(h)$ and $w''(h)$ are very 
large) has a higher density of support points. 

We mention, that the reliable and efficient computation of the Gini 
coefficient for the RJE model is a key ingredient for the numerical 
fit procedure to determine optimal parameters $a$ and $\eps$ for 
the RJE model with respect to a given raw data set.

\subsection{Duality relation for the RJE model}

The above expressions (\ref{eqFxy1}) and (\ref{eqFxy2}) provide a hint 
for a possible duality relation in the RJE model when replacing 
the auxiliary parameters of (\ref{aux}) according to 
$x\to 1/x$ and $y\to -y$. To see this, it is 
useful to express the Lorenz curve expression 
(\ref{eqsh2}), (\ref{eqRJEwofh}) in terms of the two parameters $x$ and $y$ 
which gives the nice formula:
\begin{align}
\label{eqwhxy}
w_{\rm RJE}(h)&
=\frac{W(h,x,y)}{D(a)\eps}\quad\mbox{with}\quad
W(h,x,y)=
-\frac{1-x}{y}
\ln\left(\frac{1-x\,e^{yh}}{1-x}\right)
-xh
\end{align}
where the function $W(h,x,y)$ satisfies the interesting property 
$W(h,x,y)=x\,W(h,1/x,-y)$. This means that we have the 
same Lorenz curves (and same values of the Gini coefficient) 
for two different parameter sets $(a,\eps)$ and $(a',\eps)$ 
if we can find for a given pair $(a,\eps)$ a value of $a'$ (with 
same $\eps$-value) such 
that $x(a)/D(a)=1/D(a')$, $x(a')=1/x(a)$ and $y(a')=-y(a)$. It is 
indeed possible to satisfy all three relations by choosing 
$a'=\ln(1+D(a)/x(a))=\ln(1-1/\mu(a))$. Then using the implicit 
equations (\ref{eqmu1}) and (\ref{eqmuB}) 
(reformulated for the variable $x=-D\mu$) 
one can verify that the resulting parameter values for $a'$ 
indeed satisfy $x(a')=1/x(a)$ and $y'(a)=-y(a)$ showing that the 
analytical RJE Lorenz curves 
(\ref{eqwhxy}) (and Gini coefficients as well) between $a$ and $a'$ 
(both for the same value of $\eps$ but with different $\mu$ values) 
are exactly identical. 

For example, for the data of the World 2021 Lorenz curve shown 
in Fig.~\ref{fig5}, 
we have the optimal RJE parameters $a=4.74$ and $\eps=0.0113$ corresponding 
to the blue curve in Fig.~\ref{fig5}. Using the duality relation, 
we find that there is a second solution for optimal RJE parameters 
with $a'=8.24$, $\eps'=\eps=0.0113$ which gives the identical RJE Lorenz curve 
as for $a=4.74$. This is also confirmed by the numerical code that 
determines the optimal $a$ value (by choosing a different start value for 
the iterative procedure of this code). 

\subsection{RJE Energies dependence on cumulated household}

In the RJE model (at finite $N$) the energies (corresponding to 
individual wealth) $w_m=E_m$ as a function of mode index $m$ are simply 
given by the formula $E_m=(e^{am/N}-1)/D(a)$ for $m=0,1,\ldots, N-1$. 
However, a non-trivial but very interesting quantity is the 
dependence of $E_m$ on the cumulated household $h(m)=\sum_{k=1}^m \rho_k$ 
associated to the index $m$. 
To obtain this dependence for finite $m$ and some given value of $h$ one 
has to find the specific index $m$ (function of $h$) such that 
$h(m)\le h< h(m+1)$ and take $E_{m(h)}$ (or some interpolation 
value between $E_{m(h)}$ and $E_{m(h)+1}$) for this particular 
index $m(h)$. This procedure is for finite $N$ not particularly exact 
and even for $N=10000$, we may have a significant condensation effect where 
the first values of $h(m=1)$, $h(m=2)$ are rather large (not $\sim 1/N$ 
but closer to $\sim 0.1$).

In the continuous limit $N\to\infty$ using the above results, we now have 
an easy answer for this household dependence of the RJE energy. 
We simply take the smooth function $E_s(s)=(e^{as}-1)/D(a)$, 
which is the RJE energy as a function of the continuous rescaled index 
$s=m/N$ and replace here $s\to s(h)$ given above in 
the analytic expression (\ref{eqsh2}). This gives the function:
\begin{align}
\label{eqEh1}
E_h(h)&=E_s(s(h))= 
\frac{1}{D(a)}\left(\frac{1}{e^{-as(h)}}-1\right)
=\mu\frac{1-\exp\left(\frac{a(1+\mu D(a))h}{D(a)(\eps-\mu)}\right)
}{
1+\mu D(a)\exp\left(\frac{a(1+\mu D(a))h}{D(a)(\eps-\mu)}\right)
}
\end{align}
where we have actually used the intermediate expression (\ref{eqsh1}) 
to replace $e^{-as(h)}$ which is more convenient. 

Note that we can also use the finite $N$ Lorenz curve 
construction procedure of Section 2.2  to obtain:
\begin{align}\frac{w(m+1)-w(m)}{h(m+1)-h(m)}&=
\frac{\frac{E_m}{E}\rho_m}{\rho_m}=\frac{E_m}{E}
\folgt E_h(h)=\eps w'(h)
\end{align}
taking the limit $N\to\infty$ corresponding to $h(m+1)\to h(m)$ such 
that we have the derivative $w'(h)$. This results means 
that $E_h(h)$ is related to the derivative of the Lorenz curve $w(h)$ 
(with an additional factor $E=\eps$). This can also be confirmed by a direct 
calculation of the derivative $w'(h)$ from the Lorenz curve expression 
(\ref{eqRJEwofh}) (combined with (\ref{eqsh2})) which reproduces exactly 
$E_h(h)/\eps$ with $E_h(h)$ given by (\ref{eqEh1}). 
We mention, that we have used the result $(\ref{eqEh1})$ to compute the 
RJE GDP values used in Fig.~\ref{fig13} for each country by 
$w_m^{\rm (RJE)}=E_h((m-0.5)/212)$ where $m=1,2,\ldots,212$ 
is the country index ($m=212$ for the country with maximal GDP and 
$m=1$ for the country with minimal GDP). Here, the RJE parameters 
$a$ and $\eps$ are obtained by an optimal curve fit for the Lorenz 
curve of the original GDP UN data. See also the next 
appendix section for some details on this point. 

\subsection{Cumulative distribution function}

Finally, we want to compute the cumulative distribution function 
$C(w_m)$ which gives the probability/fraction of households with an 
individual wealth larger than $w_m$. This is just $1-h(s)$ with 
$h(s)$ being the cumulated household as a function of the index variable 
$s$ given in (\ref{eqRJEhs2}) provided $s$ is determined such that 
the individual wealth/energy $w_m=E_m$ indeed corresponds to the 
index value $s$, i.e.:
\begin{align}
w_m&=E_m=E_s(s)=\frac{e^{as}-1}{D(a)}\folgt e^{as}=1+D(a)w_m
\folgt e^{-as}=\frac{1}{1+D(a)w_m}\ .
\end{align}
Inserting this expression for $e^{-as}$ in (\ref{eqRJEhs2}) we get 
(using $b=1+\mu D(a)$):
\begin{align}
C_{\rm RJE}(w_m)&
=1-\frac{(\eps-\mu)D(a)}{ab}\ln\left(\frac{1-be^{-as}}{1-b}\right)
=1-\frac{(\eps-\mu)D(a)}{ab}\ln\left(\frac{1-b\frac{1}{1+D(a)w_m}}{1-b}\right)
\\&
\label{eqCw1}
=1-\frac{\eps-\mu}{a(D^{-1}(a)+\mu )}
\ln\left(\frac{1-\mu^{-1}w_m}{1+D(a)w_m}\right)\ .
\end{align}
From this analytical result, we can deduct several limiting cases. 
First, for the RJS case $a\to 0$ (where $C(w_m)$ was not discussed in the 
first appendix section), we get:
\begin{align}
\label{eqCw0}
C_{\rm RJS}(w_m)&
=1-(\eps-\mu)\ln\left(1-\mu^{-1}w_m\right)\ .
\end{align}
Furthermore, from (\ref{eqCw1}) (and similarly for (\ref{eqCw0})), we 
obtain the linear expansion for (very) small $w_m$:
\begin{align}
C_{\rm RJE}(w_m)&\approx 
1-\frac{(\eps-\mu)( -\mu^{-1}-D(a) )}{a(D^{-1}(a)+\mu )}\,w_m
=1-\frac{(\eps-\mu)}{a}(-\mu^{-1}D(a))\,w_m
\end{align}
which is only valid for $w_m\ll \min(-\mu,D^{-1}(a))$ which is typically 
a very small range, especially for small $\eps$ with even smaller values 
of $-\mu$. This linear regime has therefore only a very limited importance. 

It is more interesting, to consider the case of a relative large value of 
$a$ (e.g. $a\sim 7$ or similar) such that $e^a\approx D(a)\gg 1$ and here 
the limit $w_m\gg \max(-\mu,D^{-1}(a))$ where typically 
$\max(-\mu,D^{-1}(a))\ll 1$. Then, we can apply in (\ref{eqCw1}) a 
different expansion in $1/w_m$:
\begin{align}
C_{\rm RJE}(w_m)&
=1-\frac{\eps-\mu}{a(D^{-1}(a)+\mu )}
\ln\left(\frac{1-\mu^{-1}w_m}{1+D(a)w_m}\right)
\\&
=1-\frac{\eps-\mu}{a(D^{-1}(a)+\mu )}
\left[-\ln(-\mu D(a))+
\ln\left(\frac{1-\mu w_m^{-1}}{1+D^{-1}(a)w_m^{-1} }\right)\right]
\\&
\approx C_1+\frac{C_2}{w_m}\quad\mbox{with}\quad
C_2=\frac{\eps-\mu}{a}
\end{align}
and
\begin{align}
\label{eqC1def}
C_1&
=1+\frac{\eps-\mu}{a(D^{-1}(a)+\mu )}\ln(-\mu D(a))
\approx \ldots 
\approx -\frac{\eps-\mu}{a}=-C_2
\end{align}
where ``$\ldots$'' represent some simplification steps 
using twice the implicit $\mu$ equation (\ref{eqmu1}) (in both ways)
and the replacement $D^{-1}(a)\approx e^{-a}$. 

In summary, we find the quite simple expression:
\begin{align}
\label{eqCw2}
C_{\rm RJE}(w_m)&\approx
\frac{\eps-\mu}{a}\left(\frac{1}{w_m}-1\right)
\end{align}
which is valid in the full interval $w_m\in[w_{\min},1]$ 
where $w_{\min}$ is some value with $w_{\min}\gg \max(-\mu,D^{-1}(a))$.

For example in the left panel of Fig.~\ref{fig20} the RJE fit 
for the Bitcoin data of 2013.Q1 (dashed blue line) is very close to 
(\ref{eqCw2}), for $w_m>3\times 10^{-5}$ and the constant term 
in (\ref{eqCw2}) 
assures the correct behavior up to $w_m\to 1$ where $C_{\rm RJE}(w_m)\to 0$. 
If we consider the smaller interval $w_m\in[3\times 10^{-5},10^{-1}]$, then 
we can also neglect the constant term in (\ref{eqCw2}) such that we 
have a pure power law $C_{\rm RJE}(w_m)\approx C_2 w_m^{-1}$ corresponding 
to the straight blue dashed line in the left panel of Fig.~\ref{fig20} 
(for this specific interval). 
In general, this pure power law regime with exponent $-1$ is only visible for 
large values of $a$, the appropriate interval for $w_m$ and small values of 
$\eps$ (i.e values of $G\sim 0.9$-$0.95$). For more modest values of $a$ or 
slightly larger values of $\eps$, we have also an intermediate regime 
with a decay $\sim \ln(1/w_m)$ (especially for the pure RJS case with 
$a=0$). 
The more refined approximation (\ref{eqCw2}) with the constant term 
(which gives a curved behavior in a double logarithmic representation) 
is in general quite accurate for a suitable 
interval $[w_{\min},1]$, even with more modest values of $a$. 

We mention, that the $C(w_m)$ curves for the RJE model (in the continuous 
limit) presented in several figures of this work are based on the 
exact expression (\ref{eqCw1}) 
or an equivalent reformulation (e.g. using (\ref{eqmu1})). 
The $C(w_m)$ curves for the finite $N$ RJE model 
(or some real data) are typically 
obtained by plotting the (rescaled) energy/individual wealth $E_m\sim w_m$ 
versus $1-h(m)$. It is also possible to reconstruct a $C(w_m)$ curve from 
a Lorenz curve but this requires good quality data and the (discrete) 
derivative of $w(h)$ to obtain $w_m\sim E_m\sim w'(h)$.

\section{Additional data for GDP of countries}

We show here additional figures for the GDP of countries and 
some additional explanations about the data shown in Fig.~\ref{fig13}. 
In Fig.~\ref{fig9}, we compare the Lorenz curves of GDP UN data of 
1973 and 2023 with theoretical Lorenz curves for the RJS and RJE models 
(with optimal values of $a$ for the RJE case). In this context, we attributed 
to each of the $212$ countries (for the 2023 data), an identical household 
fraction $\Delta h=1/212$, which can be seen by the equidistant red 
data points (with respect to the $x$-axis). However, the data points 
for the RJS/RJE models correspond to energy modes $E_k$, $k=0,1,\ldots,N-1$ 
where e.g. $N=10000$ (for the case of Fig.~\ref{fig9}) with non-uniform 
$\rho_k$ values for the household fraction for a given mode $k$ 
(also called ``agent'' with index $k$) given by the RJ formula: 
$\rho_k=T/(E_k-\mu)$. In particular for larger values of $k$ the 
value $\rho_k$ decay and the density of RJE data points on the $x$-axis 
increases. 

In this context, one can ask the question how many agents ($k$-modes) of the 
RJE model correspond to a particular country $m$ ($m=1,\ldots,212$). 
Let $h^{\rm (GDP)}(m)=m/212$ the cumulated household variable of the 
GDP raw data and $h^{\rm (RJE)}(k)=\sum_{l=0}^{k-1} \rho_l$ 
(such that $h^{\rm (RJE)}(k=N)=1$) 
the cumulated household variable for the RJE model (e.g. for the case 
with optimal fit values for $a$ and $\eps$ with respect to the GDP data).

Then, the number of agents $N_m$ associated to country $m$ is the number of 
$h^{\rm (RJE)}(k)$ values that satisfy the inequality: 
$h^{\rm (GDP)}(m-1)\le h^{\rm (RJE)}(k)<h^{\rm (GDP)}(m)$. 
This number $N_m$ typically increases with $m$ and $k$ since $\rho_k$ 
(difference between two $h^{\rm (RJE)}(k)$ values) decreases. 
Actually, at $N=10000$, the first values of $h^{\rm (RJE)}(k)$ 
($k=1,2,\ldots$) cover even several ($\sim 4-5$) countries 
(due to the effect of RJ condensation) and much larger values of $N$ 
are required to have the continuous limit. Using the analytical expression 
(\ref{eqsh2}) for the index variable $s(h)$ (dependent on the cumulated 
household $h$), we can compute the number $N_m$, or more precisely the 
fraction $f_m=N_m/N$, by $f_m=s((m+1)/212)-s(m/212)$. 

At the same time, for a given country $m$, we can also search the 
RJE energy mode $E_{k_m}$ such that $h^{\rm (RJE)}(k_m)$ 
is in the middle of the interval $[h^{\rm (GDP)}(m-1),h^{\rm (GDP)}(m)[$ 
(with best possible precision). 
This gives a virtual RJE GDP value $w_m=C E_{k_m}$ for the country $m$ 
where the constant $C$ is determined such $\sum_m w_m=1$ (sum normalized 
RJE GDP values). 
Using another analytic result (\ref{eqEh1}) of appendix A.2 for the 
RJE energy dependence $E_h(h)$ on the cumulated household $h$, we get 
directly $w_m = C E_h((m-0.5)/212)$ (for $m=1,\ldots,212$ 
and sum normalization to fix $C$). These particular effective 
RJE GDP (sum normalized) wealth values were used for the world map type 
figure Fig.~\ref{fig13}. 

\begin{figure}[htbp]
\begin{center}
\includegraphics[width=0.65\textwidth]{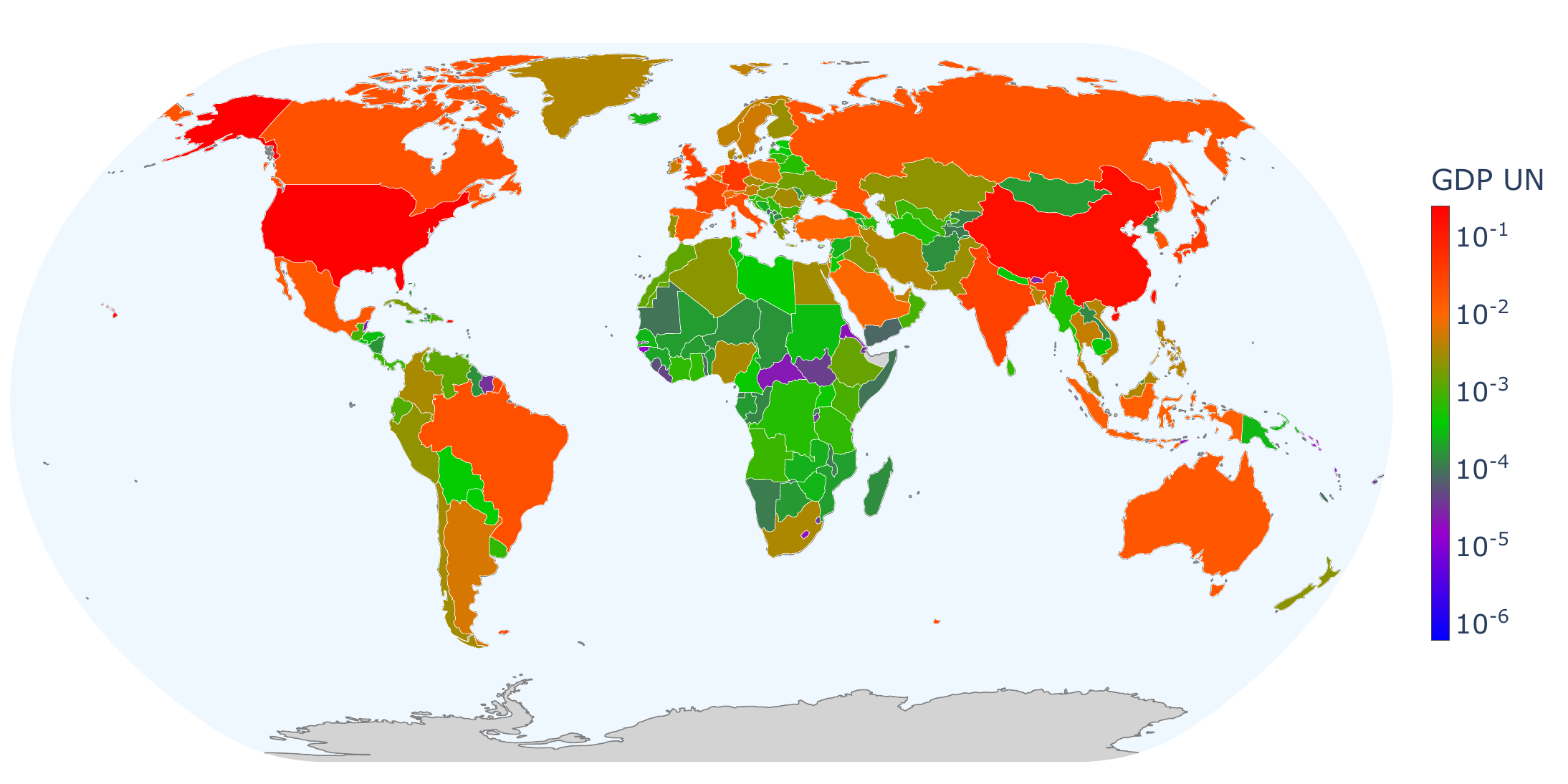}
\caption{\label{figA3}
  Same as Fig.~\ref{fig13} with
  real GDP data from UN 2023; color normalization
  is as in Fig.~\ref{fig13} 
}
\end{center}
\end{figure}

Fig.~\ref{figA3} is of the same style as Fig.~\ref{fig13} but 
it shows for comparison the (sum normalized) real GDP UN data for 2023 
(instead of the RJE GDP values). 

\begin{figure}[htbp]
\begin{center}
\includegraphics[width=0.65\textwidth]{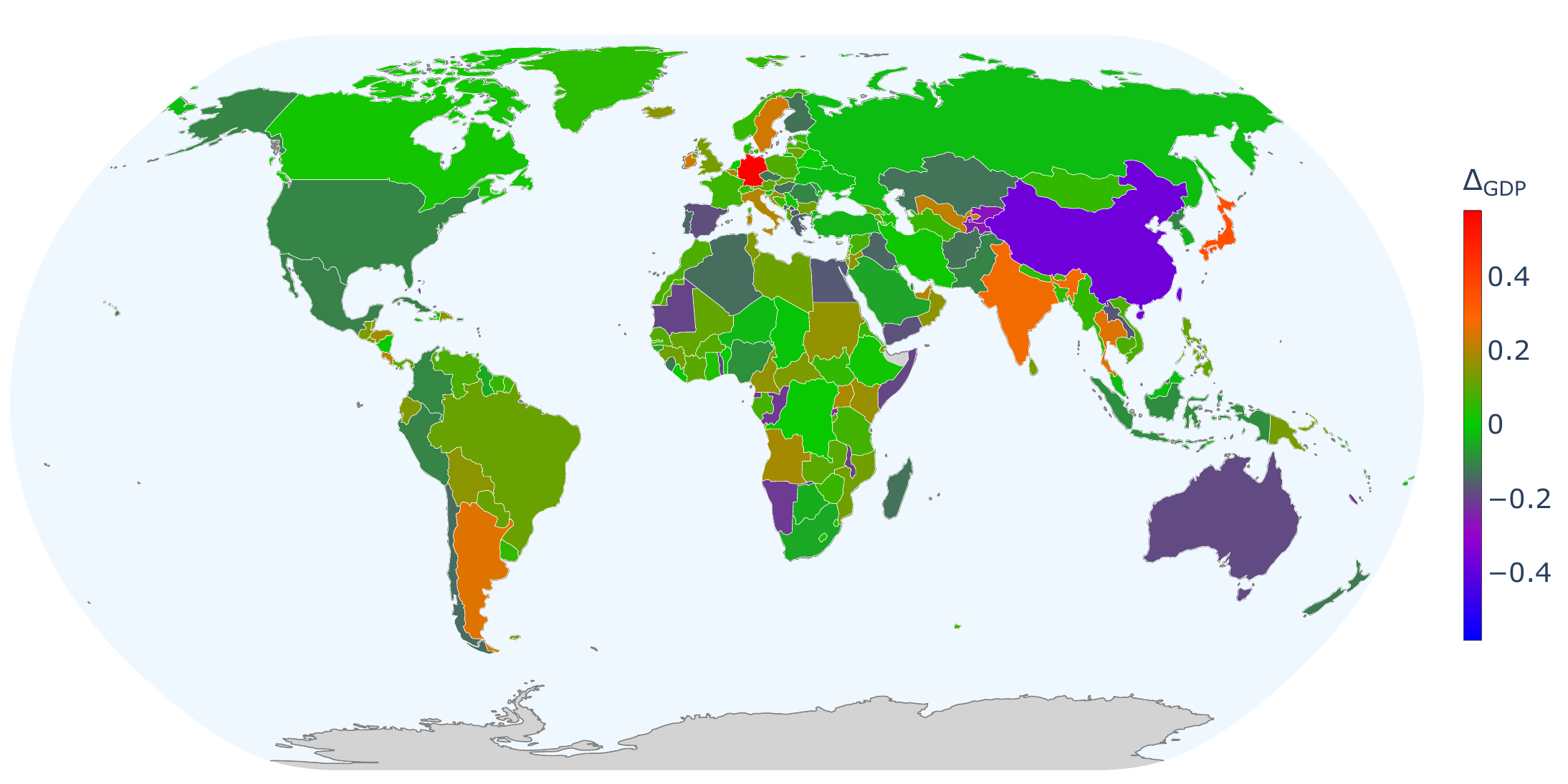}
\caption{\label{figA4}
  Color shows the relative difference $\Delta_{GDP}$ of GDP values 
  between real data (shown in Fig.~\ref{figA3}) 
  and the RJE GDP values (shown in Fig.~\ref{fig13}); here 
  $\Delta_{GDP} = \ln(w_m^{\rm (RJE)})-\ln(w_m^{\rm (UN)})
  \approx \Delta w_m/\bar w_m$ with 
  $\Delta w_m=w_m^{\rm (RJE)}-w_m^{\rm (UN)}$ 
  and $\bar w_m=(w_m^{\rm (RJE)}+w_m^{\rm (UN)})/2$. 
}
\end{center}
\end{figure}

Furthermore, Fig.~\ref{figA4} shows the relative difference between 
both GDP quantities (difference in logarithmic scale).
Examples of countries with large differences are 
Germany, Japan and India with larger RJE GDP values (by 
58\%, 35\%, 28\% respectively) and China and Australia 
with smaller RJE GPD values (by 37\%, 18\% respectively; both 
in comparison to the real UN data).

\begin{figure}[htbp]
\begin{center}
\includegraphics[width=0.65\textwidth]{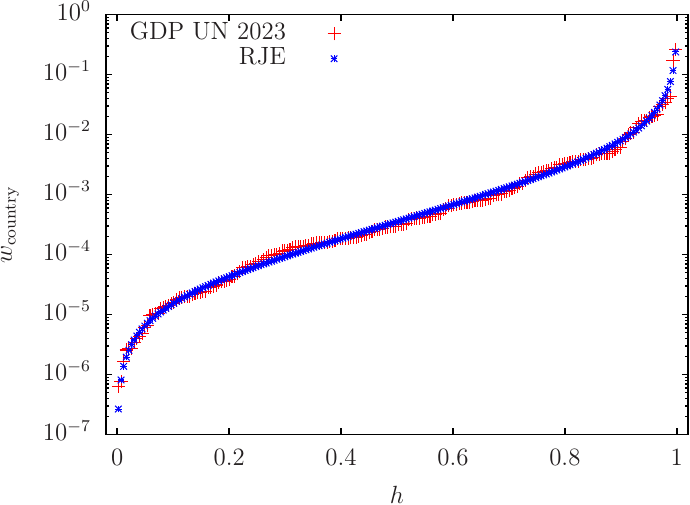}
\caption{\label{figA5}
Sum normalized country GPD wealth in logarithmic scale versus 
the rescaled country index $h=(m-0.5)/212$ for $m=1,\ldots,212$ obtained 
by ordering the world countries with increasing GDP. 
The red data points ($+$ symbols) correspond to the real GDP UN 2023 data 
and the blue data points ($*$ symbols) to the RJE GDP values 
for the optimal GDP RJE fit used 
in Fig.~\ref{fig13} (see text for details). 
}
\end{center}
\end{figure}

Fig.~\ref{figA5} shows a direct comparison of the (sum normalized) 
wealth values $w_m$ between the RJE GDP values and the UN data as a 
function of the (rescaled) country index and with a logarithmic scale 
for $w_m$. 
Both curves are close as expected but for some countries/data points there 
are significant deviations. Essentially, the RJE GDP curve in 
Fig.~\ref{figA5} corresponds to a smoothed UN data curve. Note 
that the differences of $\ln(w_m)$ between both cases correspond to the 
values used in Fig.~\ref{figA4}.

\authorcontributions{All authors equally contributed to all stages of this work.
}

\funding{The authors acknowledge support from the grant
 ANR France project
NANOX $N^\circ$ ANR-17-EURE-0009 in the framework of 
the Programme Investissements d'Avenir (project MTDINA).
This work was granted access to the HPC resources of
CALMIP (Toulouse) under the allocation 2026-P0110. 
}


{\bf Data Availability Statement:} The data presented in this study are available on request from the
corresponding author due to big amount of data.

\conflictsofinterest{The authors declare no conflict of interests.
} 



\reftitle{References}

\end{document}